\documentclass[a4paper,11pt]{article}

\usepackage[T1]{fontenc}

\usepackage{latexsym}
\usepackage{longtable}
\usepackage{epsfig}
\usepackage{graphicx,bbm,psfrag} 
\usepackage{amssymb,amsmath,amsthm,booktabs,mathtools}
\usepackage{bm}% bold math
\usepackage{color}
\usepackage{dutchcal}
\usepackage{hyperref}
\hypersetup{
    colorlinks,
    citecolor=red,
    linkcolor=blue,
}

\newtheorem{theorem}{Theorem}[section]
\newtheorem{rema}{Remark}[section]

\newtheorem{lemma}[theorem]{Lemma}

\newcommand{\bc}{\begin{center}}
\newcommand{\ec}{\end{center}}
\def\ba#1{\begin{array}{#1}\displaystyle}
\newcommand{\ea}{\end{array}}
\newcommand{\z}{\\[0.3cm] \displaystyle}
\newcommand{\beq}{\begin{equation}}
\newcommand{\eeq}{\end{equation}}
\newcommand{\beqa}{\begin{eqnarray}}
\newcommand{\eeqa}{\end{eqnarray}}
\newcommand{\no}{\nonumber}
\newcommand{\n}{\nonumber\\}
\newcommand{\bi}{\begin{itemize}}
\newcommand{\ei}{\end{itemize}}
\renewcommand{\v}[1]{\boldsymbol{#1}}

\def\micro#1{\mathrm{#1}}

\def\lt#1{\left#1}
\def\rt#1{\right#1}

\def\h#1{\hat{#1}}

\def\frc#1#2{\frac{#1}{#2}}

\newcommand{\prin}{\underline{\mathrm{P}}}
\newcommand{\p}{\partial}

\newcommand{\bra}{\langle}
\newcommand{\ket}{\rangle}
\newcommand{\dbra}{\langle\!\langle}
\newcommand{\dket}{\rangle\!\rangle}
\newcommand{\bdbra}{\Big\langle\!\!\Big\langle}
\newcommand{\bdket}{\Big\rangle\!\!\Big\rangle}
\newcommand{\Z}{{\mathbb{Z}}}

\newcommand{\R}{{\mathbb{R}}}
\newcommand{\C}{{\mathbb{C}}}
\newcommand{\Tr}{{\rm Tr}}
\newcommand{\rx}{{\rm x}}
\newcommand{\ry}{{\rm y}}
\newcommand{\rp}{{\rm p}}

\newcommand{\ep}{\epsilon}

\newcommand{\ri}{{\rm i}}
\newcommand{\re}{{\rm e}}
\newcommand{\dd}{{\rm d}}
\newcommand{\1}{{\bf 1}}
\DeclareMathOperator{\sgn}{sgn}

\newcommand{\halmos}{\rule{1ex}{1.4ex}}
\newcommand{\eproof}{\hspace*{\fill}\mbox{$\halmos$}}

\begin{document}

\begin{center}
{\Large {\sc Nonlinear projection for ballistic correlation functions:\\ a formula in terms of minimal connected covers}}

\vspace{1cm}

{\large Benjamin Doyon
}

\vspace{0.2cm}
Department of Mathematics, King’s College London, Strand, London WC2R 2LS, U.K.
\ec

\medskip

In many-body systems, the dynamics is governed, at large scales of space and time, by the hydrodynamic principle of projection onto the conserved densities admitted by the model. This is formalised as local relaxation of fluctuations in the Ballistic Macroscopic Fluctuation Theory, and is a nonlinear version of the Boltzmann-Gibbs principle. We use it to derive a projection formula, expressing $n$-point connected correlation functions (cumulants) of generic observables at different space-time points, in terms of those of conserved densities. This applies in every $d\geq 1$ spatial dimensions and under the ballistic scaling of space and time, both in and out of equilibrium. It generalises the well-known linear-response principle for 2-point functions. For higher-point functions, one needs to account for nonlinear fluctuations of conserved densities and, correspondingly, higher derivatives of local averages. Using Malyshev’s formula for the cumulant expansion, and keeping the leading order, the result is a nonlinear projection, expressed as a sum of products of correlation functions of conserved densities with equilibrium multivariances as coefficients. The sum is combinatorially organised via certain covers of the set of space-time points, which we call ``minimal connected covers''. We use this in order to get general, explicit formulas for two- and three-point functions in stationary states, expressed in terms of thermodynamic and Euler-scale data.

\tableofcontents

\section{Introduction and main results}

One of the most fundamental principles of emergent dynamical behaviours in many-body physics is that first proposed by Mori \cite{mori1965transport} and Zwanzig \cite{zwanzig1966statistical}, according to which observables project, at large scales of space and time, onto conserved densities \cite{brox1984equilibrium,spohn2012large,demasi2006mathematical,kipnis2013scaling,de2022correlation,doyon2022hydrodynamic,ampelogiannis2024long}. The principle says that there is no need to ``keep track" of all observables -- all particle trajectories, or all spin configurations, etc.~-- in order to describe what is seen at large scales of space and time: one only needs to keep track of the values of the densities of extensive conserved quantities, such as energy, momentum and particle number, in space-time. This is because on mesoscopic scales, distances much larger than microscopic distances but much smaller than the large variation scales, and in short times, conserved quantities don't change much, while arbitrary observables fluctuate a lot. Then, by ergodicity -- or more accurately, by {\em typicality} \cite{goldstein2006canonical,lanford2007entropy,allori2020statistical} -- local observables quickly take, for most times, the values they would take in the corresponding microcanonical ensemble. This holds even in systems that are not traditionally seen as ``ergodic'', such as integrable systems, which admit an infinite number of conserved densities \cite{doyon2020lecture,bastianello2022introduction,essler2023short,spohn2024hydrodynamic}. This is the hydrodynamic reduction of the dynamical degrees of freedom.

In order to illustrate this reduction, take a $d$-dimensional system\footnote{The class of dynamical systems allowed by the general theory is very wide: integrable or not as mentioned, deterministic or stochastic, classical or quantum, and with space and time that may be continuous or discrete -- in the latter case $\v x = (x_1,\ldots,x_d)$ and / or $t$ take discrete values.} with a certain number of extensive conserved quantities, $Q_i = \int \dd^d
\micro x\,q_i(\v{\micro x},\micro t),\,i=1,2,\ldots$, with associated local conserved densities $q_i(\v{\micro x},\micro t)\in\R$, $\v {\micro x}\in\R^d,\,\micro t\in\R$. For one-point functions of local observable $o(\v \rx,\micro t)$ in long-wavelength states $\bra\cdots\ket_\ell$, where $\ell$ is the variation lengthscale, one makes the hydrodynamic approximation of ``local entropy maximisation'':
\beq\label{eulerapproxintro}
	\lim_{\ell\to\infty} \bra o(\ell \v x,\ell t)\ket_\ell = \mathsf o(\mathsf q_1(\v x,t),\mathsf q_2(\v x,t),\ldots).
\eeq
Here $\mathsf o(\mathsf q_1,\mathsf q_2,\ldots) = \bra o\ket_{\mathsf q_1,\mathsf q_2,\ldots}$ is the microcanonical average of the observable $o$ on the shell determined by the values $\mathsf q_i$ of the conserved densities, and $\mathsf q_i(\v x,t)$ are the average values of these conserved densities at macroscopic coordinates $\v x,t$ in the state $\bra\cdots\ket_\ell$; these form the reduced set of dynamical degrees of freedom. Combined with conservation laws, this gives Euler hydrodynamics \cite{spohn2012large,doyon2020lecture}. Linear response -- or the Boltzmann-Gibbs principle \cite{kipnis2013scaling} -- further suggests that fluctuations of observables project onto those of conserved densities:
\beq\label{deltaolinear}
	\delta o(\v x,t)\longrightarrow
	\sum_i \frc{\p\mathsf o}{\p \mathsf q_i} \delta q_i(\v x,t).
\eeq
This is useful, as derivatives $\p\mathsf o/\p \mathsf q_i$ can be evaluated in terms of thermodynamic covariances (see e.g.~\cite{doyon2020lecture}). It gives projection formulas for two-point function \cite{spohn2012large,de2022correlation}, rigorously shown in short-range quantum spin lattices \cite{doyon2022hydrodynamic,ampelogiannis2024long}. Looking at current fluctuations $\delta \v j_i(\v \rx,t)\in\R^d$, which satisfy local conservation laws $\p_t \delta q_i + \nabla \cdot\delta \v j_i = 0$, shows that small fluctuations on top of stationary backgrounds  propagate ballistically. The associated normal modes which diagonalise these equations, and hydrodynamic velocities, are given solely in terms of thermodynamic quantities. Velocities may trivially vanish such as in purely diffusive systems, but if not, this Euler-scale theory is powerful.

But linear response describes two-point functions only. For higher-point functions, one must consider nonlinear response, a subject which has received some attention recently in non-integrable systems \cite{parker2019diagrammatic,joao2019basis,parameswaran2020asymptotically,doyon2020fluctuations,li2020nonlinear,holsten2021thermodynamic,gao2021molecular,kawabata2022nonlinear,sim2023nonlinear,fava2023divergent,delacretaz2024nonlinear} and one-dimensional integrable systems \cite{myers2020transport,fava2021hydrodynamic,de2022correlation,doyon2023emergence,doyon2023ballistic,kundu2025macroscopic,fava2024long}. Naturally, $n$-point functions are  associated to nonlinear response of $(n-1)$-th order, and thus one may expect to have to expand up to derivatives of order $n-1$ in the Taylor series expansion,
\beq\label{deltaolinear}
	\delta o(\v x,t)\longrightarrow
	\sum_i \frc{\p\mathsf o}{\p \mathsf q_i} \delta q_i(\v x,t)
	+\frc12 \sum_{ij} \frc{\p^2\mathsf o}{\p \mathsf q_i \p \mathsf q_j} \delta q_i(\v x,t) \delta q_j(\v x,t) + \ldots
\eeq
How do these higher-derivative terms combine in order to reproduce multi-point correlation functions of observables $o$? Various special cases (classes of models and / or correlations) and diagrammatic approaches have been proposed in the works mentioned above, but a simple and general formula based on simple hydrodynamic principles is still missing.

In this paper we establish this formula. It is based on the principle, or hypothesis, of {\em local relaxation of fluctuations} for fluid-cell means of (i.e.~coarse grained) observables. This principle is similar to, but stronger than, the hydrodynamic approximation \eqref{eulerapproxintro}. It says that when an observable $o$ is averaged on a mesoscopic lengthscale $L$, say on $V_L = [-L/2,L/2]^d$, and taken at macroscopically scaled coordinates $(\ell \v x,\ell t)$, the result $\overline o(\v x,t):= L^{-d} \int_{V_L} \dd^d \ry\,o(\ell \v x+\v\ry ,\ell t)$ is, for most macroscopic times $t$, well approximated by
\beq\label{meanprojintro}
	\overline o(\v x,t) \stackrel{\ell\gg L\gg\ell_{\rm micro}}\to
	\mathsf o(\overline{q_1}(\v x,t),\overline{q_2}( \v x, t),\ldots)
\eeq
where $\ell_{\rm micro}$ is a typical microscopic length scale. This is expected to hold away from shocks or other fluid singularities. This is an application of the concept of typicality to the fluid cell, and was first formulated as above in one-dimensional systems in \cite{doyon2023emergence,doyon2023ballistic}, leading to the ballistic macroscopic fluctuation theory (BMFT). It is a projection principle, where under coarse graining, an arbitrary local observable becomes a fixed function of conserved densities. It is stronger than \eqref{eulerapproxintro} because it admits that conserved densities may still fluctuate, albeit on larger scales of space and time: it states that observables' fluctuations are tied to conserved densities' fluctuations. This is, in effect, a {\em nonlinear version of the Boltzmann-Gibbs principle \cite{kipnis2013scaling}}, going beyond the central-limit scale to access large deviations. It was shown to give rise to long-range correlations in $d=1$ in \cite{doyon2023emergence}.

As we will show, in many natural states in and out of equilibrium, connected correlation functions scale in a ballistic large-deviation form, first proposed in \cite{doyon2018exact} for $d=1$:
\beq\label{scalingintro}
	\bra o_1(\ell \v x_1,\ell t_1),\ldots, o_n(\ell \v x_n,\ell t_n)\ket^{\rm c}_\ell
	\sim \ell^{(1-n)d} \,\mathsf S_{o_1,\ldots,o_n}(z_1,\ldots,z_n),
	\quad
	z_k = (\v x_k,t_k).
\eeq
This holds as a distributional equation\footnote{That is, one must take the asymptotic expansion after integrating against appropriate, fixed Schwartz functions of the scaled coordinates $\v x_1,\ldots\v x_n$, which are therefore slowly varying on microscopic scales, on both sides of \eqref{scalingintro}. An admissible class of Schwartz functions is explained in Sect.~\ref{ssecteulerscaling}. In some cases, such as in the $d=1$ hard-rod gas with a continuous distribution of velocities and for rod density observables, it holds as ordinary functions of $\v x_1,\ldots\v x_n$.} on the macroscopic coordinates $\v x_1,\ldots\v x_n$. This defines the {\em Euler amplitude} $\mathsf S_{o_1,\ldots,o_n}(z_1,\ldots,z_n)$.

The combination of Eqs.~\eqref{meanprojintro} and \eqref{scalingintro} is non-trivial. As per \eqref{scalingintro} cumulants of conserved densities have a specific asymptotic form as $\ell\to\infty$. But as per \eqref{meanprojintro}, observables are in general functions of conserved densities. Then, using a generalisation of the moment-cumulant formula to ``partial cumulants'', called Malyshev's formula \cite{malyshev1980cluster,peccati2008moments}, one may obtain the asymptotic form of cumulants of generic observables by writing them in terms of cumulants of conserved densities. This should agree with \eqref{scalingintro} for {\em every} choice of observables, and provides expressions for the Euler amplitudes of arbitrary observables, in terms of those of conserved densities.

%holds for {\em every} local observables, it says that $n$-point cumulants of any functions of the fluctuating $\overline {q_i}(z)$'s at fixed $z$ -- the microcanonical averages of local observables -- scale the same way as $n$-point cumulants of $\overline {q_{i}}(z)$'s themselves. For any probabilistic system, such a property on a group of random variables, imposes constraints on scaled cumulants of these variables, generalising Wick's theorem. These are obtained using a generalisation of the moment-cumulant formula to partial cumulants.

We will show that \eqref{meanprojintro} and \eqref{scalingintro} are indeed consistent, with the following nonlinear projection result for the Euler amplitudes.

Consider the set $Z=\{z_1,\ldots,z_n\}$. Let $\Upsilon$ be a ``minimal connected cover'' of $Z$: we define this as a cover such all patches $V\in\Upsilon$ have at least two elements, and such that, in constructing the cover patch by patch (in any order), every new patch that is adjoined to it intersects the union of all previous patches at exactly one point. To every patch $V=\{z_{k_1},\ldots,z_{k_{|V|}}\}\subset Z$, associate a connected correlation function of conserved densities,
\beq
	{\rm patch}\,V\quad \longrightarrow\quad
	\mathsf S_{i_{k_1}^V,\ldots,i_{k_{|V|}}^V}(z_{k_1},\ldots,z_{k_{|V|}})
\eeq
where for lightness of notation we use $\mathsf S_{i_1,\ldots,i_n}(z_1,\ldots,z_n):=\mathsf S_{{q_{i_1}},\ldots,q_{i_n}}(z_1,\ldots,z_n)$. This produces a certain number of ``free indices'': there are $|V|$ independent free indices, $i_{k_1}^V,\ldots,i_{k_{|V|}}^V$, for each patch $V$. Also, every ``vertex'' $z_k$ -- element of the set $Z$ -- has a certain multiplicity $m_k$: the number of patches $V$ which covers it. Then, to every vertex $z_k$, associate the $m_k$th derivative of $\mathsf o$ with respect to the $\mathsf q_{i_k^{V_r}}$'s for all patches $V_1,V_2,\ldots V_{m_k}$ that cover it, $z_k\in V_r$:
\beq
	{\rm vertex}\,z_k\quad \longrightarrow\quad
	\frc{\p^{m_k} \mathsf o}{\p \mathsf q_{i_{k}^{V_1}}\cdots \p \mathsf q_{i_{k}^{V_{m_k}}}}
\eeq
where the right-hand side is to be evaluated at $({\mathsf q}_1(z_k),{\mathsf q}_2(z_k),\ldots)$. The general projection formula is a sum over all distinct minimal connected covers, with all coefficients being unity,
\beq\label{mainintro}\boxed{
	\mathsf S_{o_1,\ldots,o_n}(z_1,\ldots,z_n) = \sum_{\Upsilon} \sum_{\rm free\,indices}\Big(\prod_{V\in\Upsilon}{\rm patch}\,V \Big)\Big(\prod_{z_k\in Z}
	{\rm vertex}\,z_k\Big).}
\eeq
This also holds as an equation for the leading asymptotics of correlation functions (again in a distributional sense), as we show that the power law from \eqref{scalingintro} adds up to $\ell^{(1-n)d}$ on the right-hand side.

This establishes the precise way in which connected correlation functions of arbitrary observables are fully determined, at leading order, by those of conserved densities. The four different types of minimal connected covers of four points are illustrated in Fig.~\ref{figcover}. It is clear from this construction that, in general, only derivatives of order up to, including, $n-1$ are required, as expected.
\begin{figure}
\bc\includegraphics[width=9cm]{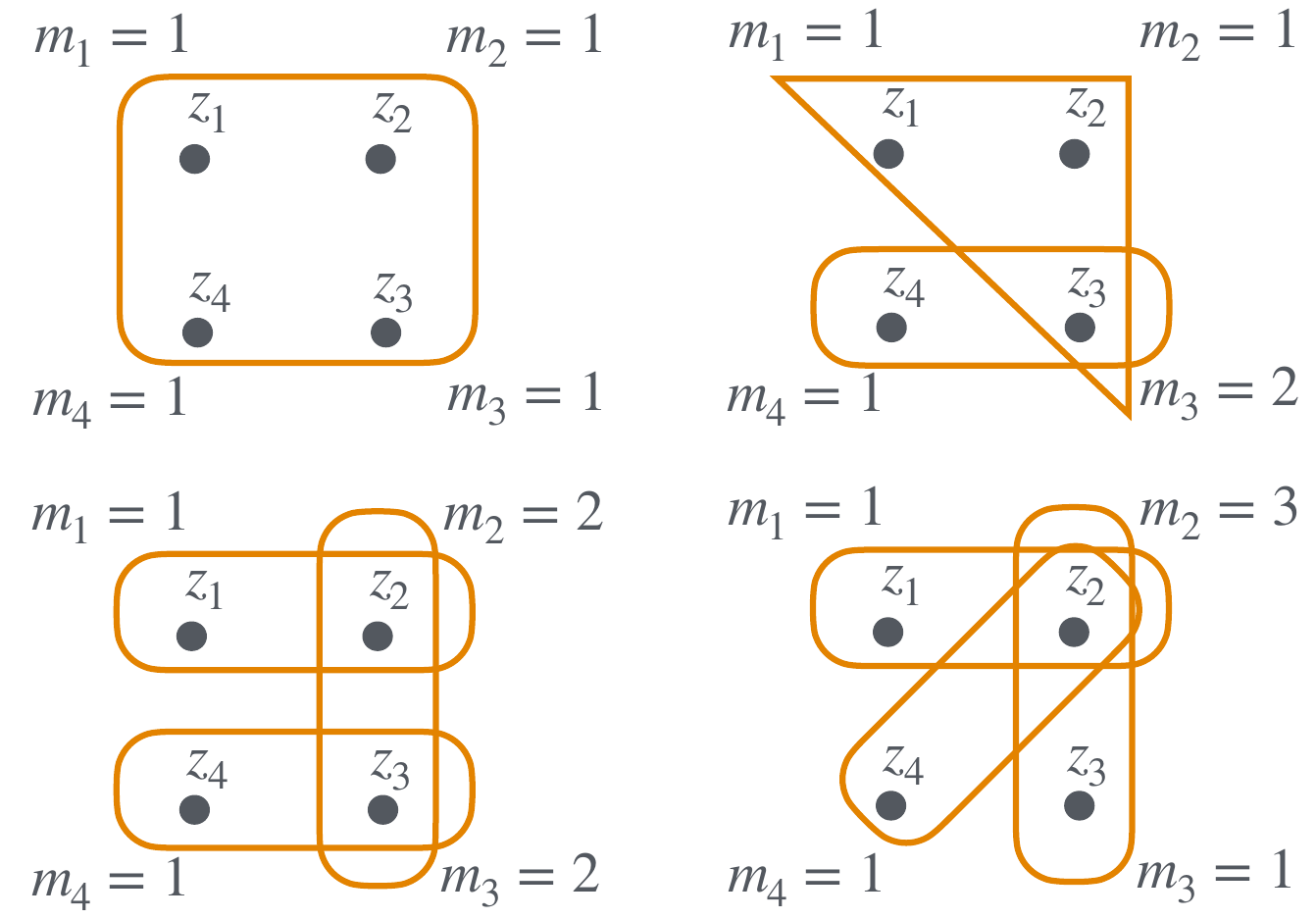}\ec
\caption{The four types of minimal connected covers for $Z=\{z_1,\ldots,z_4\}$. In each cover, the patches $V$ are the orange-colour groupings. For each type of cover, one must sum over all re-orderings of $z_k$'s leading to a different cover.}
\label{figcover}
\end{figure}
Linear response \eqref{deltaolinear} for two-point functions is reproduced, $\bra \overline{o_1}(\v x_1,t_1),\overline{o_2}(\v x_2,t_2)\ket^{\rm c} 
	\sim\sum_{ij}
	\frc{\p\mathsf o_1}{\p \mathsf q_i} \frc{\p\mathsf o_2}{\p \mathsf q_j} 
	\bra \overline{q_i}(\v x_1,t_1),\overline{q_j}(\v x_2,t_2)\ket^{\rm c}$,
%\beq\label{2ptintro}
%	\bra \overline{o_1}(\v x_1,t_1),\overline{o_2}(\v x_2,t_2)\ket^{\rm c} 
%	\sim
%	\frc{\p\mathsf o_1}{\p \mathsf q_i} \frc{\p\mathsf o_2}{\p \mathsf q_j} 
%	\bra \overline{q_i}(\v x_1,t_1),\overline{q_j}(\v x_2,t_2)\ket^{\rm c} .
%\eeq
as well as a formula for 3-point functions first obtained in \cite{doyon2023ballistic} (by the same techniques). Note that on the right-hand side of \eqref{mainintro}, we have, in general, products of distributions. But distributional parts should only occur at colliding ballistically-transported positions, as we confirm in examples. This does not cause problems because every two patches can only share at most one point.
%\beq\label{3ptintro}\begin{aligned}
%	\bra \overline{o_1},\overline{o_2},\overline{o_3}\ket^{\rm c}_\ell
%	&\sim \frc{\p \mathsf o_1}{\p \mathsf q_{i_i}}
%	\frc{\p \mathsf o_2}{\p \mathsf q_{i_2}}
%	\frc{\p \mathsf o_3}{\p \mathsf q_{i_3}}
%	\bra \overline{q_{i_1}},\overline{q_{i_2}},\overline{q_{i_3}}\ket^{\rm c}_\ell
%	\;+ \\ & \qquad + 
%	\Bigg(\frc{\p^2 \mathsf o_1}{\p \mathsf q_{i_1}\p\mathsf q_{i_1'}}
%	\frc{\p \mathsf o_2}{\p \mathsf q_{i_2}}
%	\frc{\p \mathsf o_3}{\p \mathsf q_{i_3}}
%	\bra \overline{q_{i_1}},\overline{q_{i_2}}\ket^{\rm c}_\ell
%	\bra \overline{q_{i_1'}},\overline{q_{i_3}}\ket^{\rm c}_\ell
%	+
%	\mbox{cyclic permutations $1\to 2\to 3$}
%	\Bigg)
%	\end{aligned}
%\eeq
%with Einstein's convention of summation over repeated indices, and where we kept the positions $z_k=(\v x_k,t_k)$ implicit.

For two-point function in space-time stationary states $\bra\cdots\ket$ of  quantum spin lattice systems, a formalisation of \eqref{mainintro} is shown with mathematical rigour in \cite{doyon2022hydrodynamic,ampelogiannis2024long}, using the $C^*$ algebra formulation. For two-point functions in long-wavelength non-stationary states and in dimension $d=1$, Eq.~\eqref{mainintro} was first proposed in \cite{doyon2018exact}. There, principles for the asymptotic form of higher-point functions were also proposed, and fully worked out in \cite{fava2021hydrodynamic}. But these principles were based on the assumption that local entropy maximisation holds at high enough orders in $\ell^{-1}$, which was found in \cite{doyon2023emergence} to be incorrect due to long-range correlations\footnote{For $n$-point functions where $n-1$ of the points lie in an initial state with short-range correlations, the results are still valid, and agree with \eqref{mainintro}.}. Eq.~\eqref{mainintro} finally provides the full, exact formula for all $n$-point correlation functions, and in arbitrary dimension $d\geq 1$.

The paper is organised as follows: In Section \ref{sectheorem} we express the precise mathematical theorem, Theorem \ref{theomain} proved in Appendix \ref{appproof}, at the basis of the projection formula in terms of minimal connected covers; this is a general theorem concerning algebras of observables and expectation values, independently from the setup of many-body systems. In Section \ref{sectbmft}, we introduce our many-body setup and review some basic concepts of thermodynamics and hydrodynamics, including local relaxation of fluctuations \eqref{meanprojintro} and the ballistic scaling form \eqref{scalingintro}. In Section \ref{sectgeneralprojection} we explain the main result. In Section \ref{sectapplications} we give some applications, including the exact three-point Euler amplitudes, which is a new formula in the context of hydrodynamic systems.  We conclude in Section \ref{sectconclusion}.

\section{A combinatorial theorem}\label{sectheorem}

In this section we state the general combinatorial theorem at the basis of our main result, the nonlinear projection formula. We consider a simple mathematical setup in which the theorem is proven rigorously.

Let $\alpha$ be a set, and let $\mathcal A$ be a unital, commutative, associative algebra over $\mathbb F = \R$ or $\C$ generated by $\{\1,q_a:\,a\in \alpha\}$, with unit $\1$. Let
\beq
	\dbra\cdot\dket_\ell : \mathcal A \to \mathbb F
\eeq
be a family of normalised linear maps, parametrised by $\ell>0$ and normalised to $\dbra \1 \dket_\ell = 1$. For instance, $\mathcal A$ may be an algebra of random variables in a probability space, with $\dbra\cdot\dket_\ell$ a family of expectation values. The theorem below may be generalised to non-commutative algebras; we will not need this here, but we will make some comments on this. Note that every algebra element $o$ is a term proportional to $\bf 1$ plus a polynomial in $q_{a}$'s: we denote this as $o = \mathsf o(\underline q)$,
\beq
	\mathsf o(\underline q) = c\1 + \sum_{n=1}^\infty
	\sum_{a_1,\ldots,a_n\in \alpha} c^{a_1,\ldots,a_n} q_{a_1}\cdots q_{a_n}
	\in\mathcal A,\quad c^{a_1,\ldots,a_n}\in \mathbb F
\eeq
where there are only a finite number of nonzero terms. Here and below, we use the notation
\beq
	\underline{q} = (q_{a}:a\in \alpha).
\eeq

In our particular application in the following sections, $\alpha = \mathfrak I\times \R^{d+1}$ and we write $q_a = q_i(z)$, denoting the conserved density parametrised by the index $i\in \mathfrak I$ and at space-time position $z\in \R^{d+1}$. But this choice of $\alpha$ and interpretation are not used in this section. See Sec.~\ref{sectgeneralprojection} for the connection.

Define the $n$th order cumulants
\beq
	\dbra\cdot,\ldots,\cdot\dket_\ell^{\rm c} : \mathcal A\otimes_{\mathbb F}\cdots\otimes_{\mathbb F}\mathcal A  \to \mathbb F
\eeq
in the usual way: this may be done via the moment cumulant formula
\beq\label{momentcumulantmain}
	\dbra o_1\cdots o_n\dket_\ell = \sum_{\Pi\in P(n)}\prod_{V=(j_1,j_2,\ldots,j_{|V|})\in\Pi}
	\dbra o_{j_1},\ldots,o_{j_{|V|}}\dket^{\rm c}_\ell
\eeq
where $P(n)$ is the set of partitions of $(1,2,\ldots,n)$ (in the non-commutative case, every part $V$ in the partition $\Pi$ inherits its order from $(1,2,\ldots,n)$).

The main assumption is the following. For all $n=1,2,\ldots$ and all $a_j\in \alpha$, the following limit exists:
\beq\label{Soproofq}
	\mathsf S_{a_1,\ldots,a_n} := \lim_{\ell\to\infty} \ell^{n-1}
	\dbra q_{a_1},\cdots , q_{a_n}\dket^{\rm c}_\ell.
\eeq
We are then enquiring about the similar limits, but for generic algebra elments $o_j\in\mathcal A$:
\beq\label{Soproofo}
	\mathsf S_{o_1,\ldots,o_n} := \lim_{\ell\to\infty} \ell^{n-1}
	\dbra \mathsf o_1(\underline q),\cdots ,\mathsf o_n(\underline q)\dket^{\rm c}_\ell.
\eeq
For the case $n=1$ we denote the limit
\beq\label{So1}
	\dbra \mathsf o(\underline q)\dket = S_o = \lim_{\ell\to\infty} \dbra \mathsf o(\underline q)\dket_\ell.
\eeq
Do $\mathsf S_{o_1,\ldots,o_n}$ and $\dbra \mathsf o(\underline q)\dket$ exist? If so, what values do they take?

Note that the above scaling of cumulants is associated to large-deviation theory. It appears for instance in the context of the ballistic scaling limit of correlation functions in many-body system \cite{doyon2018exact}, in the context of quantum simple exclusion processes, see the review \cite{bernard2021can}, and in diffusive hydrodynamic systems via effective field theory \cite{delacretaz2024nonlinear}. In our application, the variable $\ell$ in the factor $\ell^{n-1}$ will be replaced by $\ell^{d}$ in \eqref{Soproofq}, \eqref{Soproofo}, where $d$ is the spatial dimension (this is just a re-definition of $\ell$).

In order to answer the questions, we introduce the notion of minimal connected cover. Let us denote by $C(n)$ the set of {\em connected covers} of $\{1,\ldots,n\}$. A connected cover is a set of patches $\Upsilon = \{V_1,V_2,\ldots,V_m\}$, with $V_k\subseteq \{1,\ldots,n\}$. Each patch is a set such that $\cup_k V_k = \{1,\ldots,n\}$, and such that the {\em connectedness condition} holds: there is no subset of indices $K$ with $\cup_{i\in K} V_i \cap \cup_{i\not\in K}V_i = \emptyset$. That is, there is no subset of patches whose union is disconnected from the complementary subset of patches. For a cover $\Upsilon\in C(n)$, we denote by $m_\Upsilon(k)$ the multiplicity of the element $k\in \{1,\ldots,n\}$ within the cover $\Upsilon$, that is $m_\Upsilon(k) = |\{V\in\Upsilon: k\in V\}|$.

Let us define the ``index'' of a connected cover $\Upsilon\in C(n)$ as
\beq
	\iota(\Upsilon):=\sum_{V\in\Upsilon} |V| - |\Upsilon|.
\eeq
\begin{lemma} \label{lemmaxindex} The minimal value of the index of $\Upsilon\in C(n)$ is $\iota(\Upsilon) = n-1$.
\end{lemma}
\proof
To show that the value $n-1$ is attained we note that $\Upsilon = \{V\}=\{\{1,\ldots,n\}\}$ has index $n-1$. To show that this indeed minimises $\sum_{V\in\Upsilon} |V|-|\Upsilon|$, take a cover $\Upsilon$ with $|\Upsilon|>1$, and choose $V_1\in\Upsilon$. Because of connectedness, there must be $V_2$ that intersects $V_1$. Thus $|V_1|+|V_2| \geq |V_1\cup V_2|+1$. If $|\Upsilon|=2$ we stop there. Otherwise, again because of connectedness, there must be $V_3$ that intersects $V_1\cup V_2$. Thus $|V_1\cup V_2|+|V_3|\geq |V_1\cup V_2\cup V_3|+1$. Continuing up to $V_{|\Upsilon|}$ we find that $\sum_{V\in\Upsilon}|V| \geq n+|\Upsilon|-1$.
\eproof

We denote by $C^{\rm min}(n)\subset C(n)$ the set of {\em minimal connected covers}. This is the set of $\Upsilon\in C(n)$ that minimise its index, and that contain no singletons, $V\in\Upsilon$ with $|V|=1$:
\beq\label{defcmax}
	C^{\rm min}(n) = \{\Upsilon\in C(n):\iota(\Upsilon) = n-1,\; |V|\geq 2\,\forall\, V\in\Upsilon\}.
\eeq
Note that adding singletons to a cover does not change the index, so the condition of having no singleton makes the set $C^{\rm min}(n)$ finite. The proof of Lemma \ref{lemmaxindex} gives a method to construct all minimal connected covers, which is the one explained in Sec.~\ref{sectgeneralprojection} and in the Introduction.

We will show the following theorem.
\begin{theorem} \label{theomain}
Consider the context above, and the assumption \eqref{Soproofq}. For every $n=1,2,\ldots$: the limits \eqref{Soproofo}, \eqref{So1} exist; we have
\beq\label{onept}
	\dbra \mathsf o(\underline q)\dket = \mathsf o(\dbra \underline q\dket);
\eeq
and, for all $n\geq 2$, the following holds:
\beq\label{generalprojectionproof}
\begin{aligned}
	& \mathsf S_{o_1,\ldots,o_n}
	= \sum_{\Upsilon\in C^{\rm min}(n)}\ 
	\sum_{a_k^V\in \alpha\,:\,k\in V,V\in\Upsilon} \,\cdot\\ &\qquad\qquad\qquad
	\Bigg(\prod_{\{k_1,\ldots,k_{|V|}\}=V\in\Upsilon} \mathsf S_{a^V_{k_1},\ldots,a^V_{k_{|V|}}}\Bigg)\,\times\,
	\Bigg(
	\prod_{k=1}^n \frc{\p^{m_\Upsilon(k)}\mathsf o_k(\underline{\mathsf q})}{\prod_{V\in\Upsilon_k}\p \mathsf q_{a^V_k}}\Bigg|_{\underline{\mathsf q} = \dbra \underline{q}\dket}\Bigg).
	\end{aligned}
\eeq
\end{theorem}
The proof of this theorem is provided in Appendix \ref{appproof}. Although the theorem has a distinctive ``large-deviation'' flavour (see the next section), the proof is combinatorial. We are not aware of the statement of the above theorem in the literature.

\section{Principles of hydrodynamics}\label{sectbmft}

In this section, we review aspects of the thermodynamics and Euler-scale hydrodynamics of general many-body systems in spatial dimension $d$. This includes the generalisation to dimensions $d>1$ of some of the concepts introduced recently: local relaxation of fluctuations \cite{doyon2023ballistic}, the ballistic large-deviation scaling of correlation functions \cite{doyon2018exact}, and the reduced measure on coarse-grained conserved densities \cite{doyon2023ballistic}.

\subsection{Conserved quantities and ensembles}\label{secconsqties}

Consider a dynamical system in $d$ spatial dimensions. We wish to be as general as possible about the system, and thus loosely specify our setup as follows.

The system admits an algebra of local observables. For instance, these are functions on phase space in Hamiltonian systems, random variables of the configurations in stochastic systems, operators on the quantum Hilbert space or more generally elements of some operator algebra, etc. As local observables form an algebra, any finite product of local observables is also a local observable. There is an abelian group of space-translation automorphisms: for every local observable $o$, it acts as $o\mapsto o(\v\rx)$, where $\v\rx=(\rx_\mu)_\mu = (\rx_1,\ldots,\rx_d)\in\R^d$ is the spatial position. As is habitual in this setup, space may indeed be continuous as in quantum or classical gases or field theories, or may be discrete as in quantum or classical chains or lattice models. In the latter case we replace $\R^d$ by, say, $\Z^d$, and define $o(\v\rx) =o(\lfloor\v\rx\rfloor) =  o(\lfloor \rx_1\rfloor,\ldots,\lfloor \rx_d\rfloor)$; but this will not play any role here, as we analyse the system at large scales, where space and time are continuous. Further, there is a notion of spatial support of an observable: $o(\v\rx)$ may have finite support around $\v\rx$ (strictly local), or infinite support with appropriately decaying ``norm'' away from the position $\v\rx$ (quasi-local); below we use ``local'' to mean local or quasi-local. The dynamics is obtained as a one-dimensional group of transformations of observables, $o(\v\rx)\to o(\v\rx,\micro t)$. Time evolution can be Hamiltonian or not, and we assume that  it is an automorphism of the algebra, and that the system is homogeneous: the dynamics is space-translation invariant (i.e.~space- and time-translations commute).

Most importantly, the system admits a certain number of extensive conserved quantities
\beq
	Q_i = \int \dd^d \rx\,q_i(\v\rx,\micro t) = \int_{\R^d} \dd^d \rx\,q_i(\v\rx,\micro t),\quad
	\p_{\micro t} Q_i = 0
\eeq
where $\v\rx=(\rx_1,\ldots,\rx_d)\in\R^d$ is the spatial position and $q_i(\v\rx,\micro t)$ is the associated local density (in discrete time, $\p_{\micro t}$ is discrete partial derivative). The index $i$ lies in some index set $\mathfrak I$ which may be finite or not; the index may be discrete or continuous but we will use the notation $\sum_i$ to indicate a sum or integral over all conserved quantities. We assume that there are associated local currents $\v j_i(\v\rx,\micro t) = (j_{i,1}(\v\rx,\micro t),\ldots,j_{i,d}(\v\rx,\micro t))$ giving rise to local conservation laws:
\beq\label{cons}
	\p_t q_i(\v \rx,\micro t) + \nabla\cdot \v j_i(\v\rx, \micro t) = 0
\eeq
where $\nabla = (\p_{\rx_1},\ldots,\p_{\rx_d})$ (or its discrete version for systems on discrete space).

We recall the macrocanonical ensemble: these are the maximal entropy states under the constraints of the extensive conserved quantities,
\beq\label{macrodef}
	\bra \cdots\ket_{\underline\beta}
	=\lim_{\mathcal V\to\R^d} \frc{\int \dd\mu \,\re^{-\sum_{i} \beta^i Q_i}
	\cdots}{\int \dd\mu \,\re^{-\sum_{i} \beta^i Q_i}}.
\eeq
Here $\underline\beta = (\beta^1,\beta^2,\ldots)$ are the ``Lagrange parameters'', now the charges $Q_i = \int_{\mathcal V} \dd^d \rx\,q_i(\v\rx)$ lie on the volume $\mathcal V$, and the large-volume limit $\mathcal V\to\R^d$ is taken and assumed to exist for averages of local observables. The symbol $\mu$ stands for a time-invariant, homogeneous prior measure. It is difficult to define this prior measure universally, but, for instance, in classical Hamiltonian systems of particles with canonical coordinates $(\v\rx_a,\v\rp_a)$, one may choose $\dd\mu = \sum_{N=0}^\infty (N!)^{-1} \prod_{a=1}^N \dd^d \rx_a \dd^d \rp_a$ (direct sum); while in quantum systems, $\bra \cdots\ket_{\underline\beta}=\frc{\Tr \,\re^{-\sum_{i} \beta^i Q_i}\cdots}{\Tr \,\re^{-\sum_{i} \beta^i Q_i}}$ with the trace on an appropriate Hilbert space. We concentrate on regions of the parameters $\underline\beta$ where the limit \eqref{macrodef} does not depend on the boundary conditions on $\p\mathcal V$, and averages of local observables are smooth in $\underline\beta$, thus avoiding phase transitions. Likewise, the microcanonical ensemble is
\beq
	\bra \cdots\ket_{\underline{\mathsf q}}
	=\lim_{\mathcal V\to\R^d} \frc{\int_{Q_i/|\mathcal V|\,\in\,[\mathsf q_i-\ep,\mathsf q_i+\ep]\,\forall i} \dd\mu \,
	\cdots}{\int_{Q_i/|\mathcal V|\,\in\,[\mathsf q_i-\ep,\mathsf q_i+\ep]\,\forall i} \dd\mu}
\eeq
where $\ep\to0$ as $\mathcal V\to\R^d$ in an appropriate fashion \cite{aizenman1978conditional,brandao2015equivalence}. Assuming strict positivity of the covariance matrix (it is clearly non-negative, but one needs to assume strict positivity, which is equivalent to convexity of the free energy)
\beq\label{Cmatrix}
	\mathsf C_{ij} = -\frc{\p\bra q_j\ket_{\underline\beta}}{\p \beta^i} = \bra Q_i,q_j(\v 0)\ket^{\rm c}_{\underline\beta} = \int \dd^d\rx \,\bra q_i(\v\rx),q_j(\v 0)\ket_{\underline\beta}^{\rm c} =
	\int \dd^d\rx \,\big(\bra q_i(\v\rx)q_j(\v 0)\ket_{\underline\beta}
	-
	\bra q_i(\v\rx)\ket_{\underline\beta}\bra q_j(\v 0)\ket_{\underline\beta}\big)
\eeq
with
\beq
	\mathsf C^{\rm T} = \mathsf C,\quad \mathsf C>0,
\eeq
the map $\underline\beta\mapsto\bra\underline q\ket_{\underline\beta}$ is invertible. This defines functions $\beta^i(\underline{\mathsf q}) = \beta^i(\mathsf q_1,\mathsf q_2,\ldots)$ for all $i$'s. There is equivalence between microcanonical and macrocanonical ensembles \cite{aizenman1978conditional,touchette2015equivalence,brandao2015equivalence}: for every local obsevable $o$, we have $\bra o\ket_{\underline{\mathsf q}} = \bra o\ket_{\underline\beta(\underline{\mathsf q})}= \bra o(\v \rx,\micro t)\ket_{\underline{\mathsf q}} = \bra o(\v \rx,\micro t)\ket_{\underline\beta(\underline{\mathsf q})}$. We denote the resulting average as a function of conserved densities $\underline{\mathsf q}$, by using the ``sans-serif'' font for the observable,
\beq\label{defo}
	\mathsf o(\underline{\mathsf q}) := \bra o\ket_{\underline{\mathsf q}} = \bra o\ket_{\underline\beta(\underline{\mathsf q})}.
\eeq

We recall that derivatives with respect to conserved densities are related to covariances in the {\em macrocanonical} ensemble. For instance, by the chain rule, we find\footnote{Here and below, sums on conserved density indices are implicitly over $\mathfrak I$.}
\beq\label{deroq}
	\frc{\p\mathsf o}{\p \mathsf q_i}
	= \sum_j \mathsf C^{ij} \bra Q_j,o\ket_{\underline \beta}^{\rm c} 
\eeq
where $\mathsf C^{ij}$ is the inverse of the covariance matrix $\mathsf C_{ij}$, that is $\sum_j\mathsf C_{ij}\mathsf C^{jk} = \delta_i^{~k}$,  and likewise
\beq
	\frc{\p^2\mathsf o}{\p \mathsf q_i \p \mathsf q_j}
	= \sum_{kl} \mathsf C^{ik}\mathsf C^{jl} \bra Q_k,Q_l,o^-\ket_{\underline \beta}^{\rm c}
\eeq
where
\beq\label{ominus}
	o^- := o - \sum_{ij} q_i \mathsf C^{ij} \bra Q_j,o\ket_{\underline \beta}^{\rm c}
\eeq
is the observable $o$ from which its overlap with the conserved quantities has been projected out.

Further properties of such thermodynamic quantities are explained in App.~\ref{appmatrices}.

\subsection{Fluid-cell mean and local relaxation of fluctuations}\label{ssectrelax}

We now discuss the main tool for the present paper: the principle (hypothesis) of local relaxation of fluctuations. This was introduced in the context of the BMFT in one-dimensional systems \cite{doyon2023emergence,doyon2023ballistic}, but it is more general; it is a particular implementation of the hydrodynamic projection of Mori-Swanzig \cite{mori1965transport,zwanzig1966statistical}. This is the principle according to which arbitrary observables averaged over ``fluid cells'' project, in a sense, onto conserved densities. Here we present an intuitive definition of the fluid-cell mean, which makes our derivation clear; however more formal expressions, in terms of Fourier transform, may be more appropriate for rigorous developments, see e.g.~\cite{doyon2022hydrodynamic,ampelogiannis2024long}.

The main formula is \eqref{meanproj}. But first, let us provide heuristic arguments to explain how to get there.

Let $L$ to be a mesoscopic length, much larger than microscopic lengths $\ell_{\rm micro}$ (typical distances between particles, typical interaction ranges, etc.), but much smaller than $\ell$, the macroscopic variation length scale. Consider a region of linear size $L$; the precise shape is not important, but one may take the box
\beq
	V_L = [-L/2,L/2]^d.
\eeq

On this region, at time $\micro t$, consider the mean of a conserved density,
\[
	L^{-d}\int_{V_L} \dd^d\rx\,q_i(\v\rx,\micro t).
\]
This is a ``macrovariable'', on this mesoscopic scale. Because of the conservation law, the time derivative of this is $-L^{-d} \int_{\p V_L} \dd^{d-1} \v \rx\cdot \v j_i(\rx,t)$ where $\dd^{d-1} \v \rx$ is the surface measure vector outwardly perpendicular to the surface. This fluctuates as time goes forward, for generic initial configuration. But fluctuations only occur because of interactions at the surface $\p V_L$. Assuming a finite density, and that interaction ranges are on the (finite, fixed) microscopic scale $\ell_{\rm micro}$, such microscopic fluctuations occur because of ``interaction events'' (such as collisions of spheres in the hard-sphere gas) within the finite-width shell around the cell's surface of dimension $L^{d-1}$. Hence on very short times $\tau\sim L^{1-d}$ we expect that the fluid-cell mean of the conserved quantity does not change. This macrovariable can be considered fixed, non-fluctuating: it is a constraint.

Now consider the mean of an observable that is {\em not conserved}:
\[
	L^{-d}\int_{V_L} \dd^d\rx\,o(\v\rx,\micro t).
\]
This is also a macrovariable, and also fluctuates; however because it is not conserved, changes to this quantity occur many more times on any time interval $\tau$. Indeed, they occur, loosely speaking, whenever there is some interaction between the microscopic degrees of freedom (such as particles) within the region $V_L$. Again at finite densities and with interaction ranges of order $\ell_{\rm micro}$, we expect $\mathcal O(L^d)\times \tau$ independent fluctuations to occur in time $\tau$ (for instance, again, collisions between hard spheres in a finite-density hard-sphere gas).
%We expect that $\mathcal O(L^d)$ microscopic interaction events -- whereby almost every finite-volume patch within the cell is ``affected'' a finite number of times -- carry any information from changes at the boundary of the cell, a finite distance within the cell. 
With $\tau = \mathcal O(L^{1-d})$, there are $\mathcal O(L)$ fluctuations. Hence on this very short time, when the mean conserved densities are constants, the fluid-cell mean of the observable $o$ has fluctuated much more. One can expect, by a ``molecular chaos'' principle, that the re-organisation of the cell's configuration over time $\tau$ allows the mean observable to continuously adjust to the slowly varying mean conserved density, and relax to its microcanonical value characterised by these mean conserved densities. Of course $\mathcal O(L)$ fluctuations in a volume where there are $\mathcal O(L^d)$ degrees of freedom is not sufficient in general for relaxation. But over time, locally the system relaxes, so all that is required is that the $\mathcal O(L)$ times more fluctuations undergone by the mean obsevable $o$ as compared to the mean conserved density, allows it to follow it and keep its microcanonical value.

Here we implicitly used {\em typicality} (see e.g.~\cite{goldstein2006canonical,lanford2007entropy,allori2020statistical}): there is no need for time averaging, as for most times, the average over space takes the typical value.

Defining the {\em fluid-cell mean} of an observable $o$ at macroscopic coordinates $\v x,t$ as
\beq
	\overline o(\v x,t):= \frc1{L^d} \int_{V_L} \dd^d \ry\,
	o(\ell \v x+\v\ry ,\ell t),
\eeq
these arguments suggest that (see Eq.~\eqref{meanprojintro}),
\beq\label{meanproj}
	\overline o(\v x,t) \stackrel{\ell\gg L\gg\ell_{\rm micro}}\to
	\mathsf o(\overline{\underline q}(\v x, t))
	= \mathsf o(\overline{q_1}(\v x,t),\overline{q_2}( \v x, t),\ldots)
\eeq
This is the hypothesis of {\em local relaxation of fluctuations}.

\begin{rema}\label{remashocks} Eq.~\eqref{meanproj} is expected to hold ``almost everywhere'' in space-tiome. It assumes that, on the mesoscopic scale $L$, the many-body system is in a homogeneous state. This breaks if there is a fluid singularity at $(\v x,t)$, such as a shock, or if this limit is taken within a correlation with another local observable at the same space-time point.
\end{rema}

\subsection{Long-wavelength states and Euler-scale equations}

In this section, we specify the type of states that we consider, and recall how the Euler-scale equations of the system can be derived. The Euler-scale equation will play an important role in our derivation of the scaling of correlation functions below, and this will also serve to illustrate how the principle of local relaxation of fluctuations can be applied.

\subsubsection*{Long-wavelength states}

The principle of Euler hydrodynamics are most easily illustrated by considering some long-wavelength state, that is not necessarily locally stationary. One may take
\beq\label{lw}
	\bra \cdots\ket_\ell
	= \frc{\int \dd\mu \,\re^{-\sum_{o\in S} \int\dd^d \rx \,\beta^o(\v\rx/\ell) o(\v\rx)}
	\cdots}{\int \dd\mu \,\re^{-\sum_{o\in S} \int\dd^d\rx \,\beta^o(\v\rx/\ell) o(\v\rx)}}
\eeq
for some set $S$ of local observables, associated set $\{\beta^o(\cdot):o\in S\}$ of space-dependent conjugate fields, and large, macroscopic scale $\ell$. Typically, one takes the set $S$ to be composed of local conserved densities, $S = \{q_1,q_2,\ldots\}$, in which case the initial state is locally stationary; but this is not necessary for the general discussion. This state is locally translation invariant, only showing $O(1)$ variations of local averages on distances of order $\ell$. It is expected to have short range correlations in general at least for high enough ``temperatures'' $\beta^o(\v x)$ (this is rigorously shown, for instance, in quantum lattice systems \cite{bratteli1987operator,bratteli1997operator}). These are just a simplifying features, not essential for our resuls. For instance, states with long-range correlations may be represented by adding multilinear terms in the exponential, see \cite{PhysRevLett.134.187101} (these terms develop over time from inhomogeneous states such as \eqref{lw}, according to BMFT results \cite{doyon2023emergence} -- so long-range correlations are generically present at all macroscopic times $t>0$).

\subsubsection*{The Euler hydrodynamic equation from local relaxation of fluctuations}

The question is to establish an equation for the quantities
\beq\label{mathsfqi}
	\mathsf q_i(\v x,t) := \lim_{\ell\to\infty} \bra q_i(\ell \v x,\ell t)\ket_\ell.
\eeq
The main result of this paper, and in particular the higher-point functions studied in Sec.~\ref{sectapplications}, will be a generalisation of this to higher-point functions, so it is useful to illustrate some of the ideas here.

For this purpose, one may simply combine the hydrodynamic approximation of local entropy maximisation \eqref{eulerapproxintro}, with the local conservation laws \eqref{cons}, to get (see for instance \cite{doyon2020lecture} for how this is done -- it is explained in $d=1$ but easily adapted to $d>1$)
\beq\label{eulerhydro}
	\p_t \mathsf q_i(\v x,t) + \nabla\cdot \v{\mathsf j}_i(\underline{\mathsf q}(\v x,t)) = 0
\eeq
where $\v{\mathsf j}_i(\underline {\mathsf q}) = (\mathsf j_{i,1}(\underline {\mathsf q}),\ldots,\mathsf j_{i,d}(\underline {\mathsf q}))$ and each $\mathsf j_{i,r}(\underline{\mathsf q})$ is defined in \eqref{defo}. Now $\p_t$ and $\nabla$ are continuous derivatives, as the macroscopic coordinates $t\in\R$ and $\v x\in\R^d$ are continuous variables.

But local relaxation of fluctuations \eqref{meanproj} is perhaps better justified than the ad-hoc local entropy maximisation \eqref{eulerapproxintro}, and gives more insight.

First, it implies that {\em fluid-cell means of conserved densities are transported according to the Euler-scale equation}  (as first proposed in \cite{doyon2023emergence,doyon2023ballistic} in $d=1$). Although we will not use this here, we explain it as an illustration. Observe that the local conservation laws \eqref{cons} hold for the fluid-cell means, $\p_{ t} \overline{q_i}(\v x, t) + \nabla\cdot \overline{\v j_i}(\v{x}, t) = 0$. Therefore, using \eqref{meanproj}, one can derive
\beq\label{eulermean}
	\p_{ t} \overline{q_i}(\v x, t) + 
	\nabla \cdot \v{\mathsf j}_i(\overline{\underline{q}}(\v x, t)) = 0\qquad (\ell\gg L\gg\ell_{\rm micro}).
\eeq
Here, the spatial derivatives must be {\em macroscopic derivatives}, while the time derivative may be microscopic. That is, we may replace $\p_t \overline{q_i}(\v x,t)$ by $\ell\overline{\p_{\micro t} q_i}(\v x,t)$, but we cannot do the equivalent for the spatial derivatives.

The derivation goes as follows. We first write, from the conservation laws,
\beqa\label{transportfluctu}
	\lefteqn{\ep^{-d} \int_{[0,\ep]^d} \dd^d y\, \p_t \overline{q_i}(\v x + \v y,t)} &&\\
	&& +\ \ep^{-(d-1)} \int_{[0,\ep]^{d-1}} \dd^{d-1} y\, \ep^{-1} \sum_r \big(\overline{j_{i,r}}(\v{x}+\v\ep_r+\v y_{(1,\ldots,d)\setminus r},t)  - \overline{j_{i,r}}(\v{x}+\v y_{(1,\ldots,d)\setminus r}, t)\big) = 0\no
\eeqa
where $\v\ep_r = (0,\ldots,0,\underbrace{\ep}_{\text{position $r$}},0,\ldots,0)$
and $\v y_{(1,\ldots,d)\setminus r}= (y_1,\ldots,y_{r-1},\underbrace{0}_{\text{position $r$}},y_{r+1},\ldots,y_{d-1})$. Then, we derive \eqref{eulermean} by taking the limit $\lim_{\ep\to 0}\lim_{\ell\gg L\gg\ell_{\rm micro}}$, in that order. In doing so, $\overline{j_{i,r}}$ may be first replaced by its relaxed value according to \eqref{meanproj}, and the second term in \eqref{transportfluctu} indeed gives the macroscopic spatial derivatives of \eqref{eulermean}.

Note that in general, {\em $\overline{q_i}(\v x,t)$'s are themselves fluctuating} in the state \eqref{lw}, hence \eqref{eulermean} is a transport equation for fluctuating variables. This is the basis for BMFT: transporting initial fluctuations through space-time. Euler-scale equations typically develop singularities in finite time, and weak solutions are not unique. As the initial fluctuations in the state \eqref{lw} make coarse-grained densities $\overline{q_i}(\v x,t)$ rough, Eq.~\eqref{eulermean} does not define uniquely how to transport initial fluctuations. But this is avoided in linearly degenerate systems \cite{lax2005hyperbolic,ferapontov1991integration,el2011kinetic,pavlov2012generalized} (see the lecture notes \cite[Sec 3.2]{bressan2013hyperbolic}), as discussed in \cite{doyon2023ballistic}. Here, we will not directly use transport of fluctuations, Eq.~\eqref{transportfluctu}, thus we avoid these subtleties.

Second, {\em local relaxation of fluctuations \eqref{meanproj} also implies local entropy maximisation, Eq.~\eqref{eulerapproxintro}}, without the need for the fluctuation-transport equation \eqref{transportfluctu}. This happens if we further assume that {\em the cumulant matrix of fluid-cell means of conserved densities vanish}:
\beq\label{vanish2pt}
	\bra \overline{q_i}(\v x,t),\overline{q_j}(\v x,t)\ket_\ell^{\rm c} 
	\stackrel{\ell\gg L\gg\ell_{\rm micro}}\to 0.
\eeq
Indeed, by \eqref{meanproj}, we have
\beq\label{obarproj}
	\bra \overline o(\v x,t)\ket_\ell \sim
	\bra \mathsf o(\overline{\underline{q}}(\v x,t))\ket_\ell\quad (\ell\gg L\gg\ell_{\rm micro})
\eeq
and by the Bienaym\'e-Chebyshev inequality, \eqref{vanish2pt} implies that we may replace the observable $\overline{q_i}(\v x,t)$ by its average $\bra \overline {q_i}(\v x,t)\ket_\ell=\mathsf q_i(\v x, t)$ on the right-hand side of \eqref{obarproj}, thus obtaining \eqref{eulerapproxintro} and therefore \eqref{eulerhydro}.

The BMFT results \cite{doyon2023ballistic,doyon2023emergence} (valid in one-dimensional systems with the property of linear degeneracy) guarantee that, although long-range correlations develop, these are not strong enough and \eqref{vanish2pt} holds at all times $t$ for initial states such as \eqref{lw}. In fact, Eq.~\eqref{vanish2pt} is expected to hold very generally: it simply indicates that the cumulants of total conserved quantities on regions of volume $L^d$ vanish more quickly than $L^{2d}$:
\beq
	\int_{V_L} \dd^d\ry \int_{V_L} \dd^d\ry' \bra q_i(\ell x+\ry,\ell t) q_j(\ell x+\ry',\ell t)\ket^{\rm c}_\ell \ll L^{2d}
\eeq
which is the case, by the bounded convergence theorem, as soon as correlations decay in space, $\bra q_i(\ell x+\ry,\ell t) q_j(\ell x+\ry',\ell t)\ket^{\rm c}_\ell\to 0$ for all $|\ry|+|\ry'|= \alpha L$ and $\alpha>0$.

Note how {\em very little of the precise form of the initial state \eqref{lw} is used, and there is no need to assume any strong version of local entropy maximisation}: the state at $\v x,t$ does not need to have, locally, just the correlations of a maximal entropy state. This is important, because by the BMFT \cite{doyon2023emergence}, and its expected generalisation to dimensions $d$, spatial correlations of order $1/\ell^d$ develop. The local state only needs to reproduce local averages in the strict limit $\ell\to\infty$, which happens whenever {\em local relaxation of fluctuations hold and two-point correlation functions of local densities vanish, no matter how slowly, at large distances}.

\subsection{Ballistic scaling of connected correlation functions}\label{ssecteulerscaling}

Let us now derive the following large-deviation, ballistic scaling of connected correlations functions of local observables. This will play a crucial role in our derivation of the main formula \eqref{mainintro}. In order to ligthen the notation, we will sometimes use
\beq
	z_k = (\v x_k,t_k),\quad z=(\v x,t)\qquad \in \R^{d+1}.
\eeq

The large-deviation scaling of connected correlation functions was first proposed in \cite{doyon2018exact} in $d=1$, and is here extended to higher dimensions: we will show that the asymptotic expansion has the form
\beq\label{scalingo}
	\bra \overline {o_1}( z_1),\ldots, \overline {o_n}( z_n)\ket^{\rm c}_\ell \sim \ell^{(1-n)d}\,\mathsf S_{o_1,\ldots,o_n}(z_1,\ldots,z_n)
\eeq
as $\ell\gg L\gg \ell_{\rm micro}$. Here, $\bra \cdot,\ldots,\cdot\ket^{\rm c}_\ell$ is the connected correlation function within the state $\bra \cdots\ket_\ell$ of the form \eqref{lw}, and the above defines the {\em Euler amplitudes} $S_{o_1,\ldots,o_n}(z_1,\ldots,z_n)$ (independent of $\ell$ and $L$). We will also use the simplified notation
\beq
	\mathsf S_{i_1,\ldots,i_n}(z_1,\cdots,z_n)
	:=
	\mathsf S_{q_{i_1},\ldots,q_{i_n}}(z_1,\cdots,z_n).
\eeq

Two comments are in order, in order to specify in what sense \eqref{scalingo} is expected to hold.

\subsubsection*{Eq.~\eqref{scalingo} as a distributional equation}

The first comment is that Eq.~\eqref{scalingo} holds {\em as a distributional equation on the spatial coordinates}: the Euler amplitudes are, generically, distributions. That is, the large-$\ell$ asymptotics holds after integration against Schwartz functions of the scaled spatial coordinates $f(\v x_1,\ldots,\v x_n)$. Therefore, in fact, the fluid-cell averaging can be discarded: once we integrate against Schwartz functions, this corresponds to changing the function to $f(\ldots,\v x_i+\ry/\ell,\ldots)$ and averaging over $|\ry|\leq L/2\ll \ell$, which gives the same function in the limit. That is
\beq\begin{aligned}
	& \lim_{\ell\to\infty} \ell^{(n-1)d}\,\int \dd^dx_1 \cdots \dd^dx_n \,f(\v x_1,\ldots,\v x_n) \bra \overline {o_1}( \v x_1, t_1),\ldots, \overline {o_n}( \v x_n, t_n)\ket^{\rm c}_\ell\\ = &
	\lim_{\ell\to\infty} \ell^{-d}\,\int \dd^d\rx_1 \cdots \dd^d\rx_n \,f(\v \rx_1/\ell,\ldots,\v \rx_n/\ell) \bra  {o_1}( \v \rx_1, \ell t_1),\ldots,  {o_n}( \v \rx_n, \ell t_n)\ket^{\rm c}_\ell
	\\ & \qquad\qquad\qquad\qquad = \int \dd^dx_1 \cdots \dd^dx_n \,f(\v x_1,\ldots,\v x_n)\mathsf S_{o_1,\ldots,o_n}(\v x_1,t_1;\cdots;\v x_n,t_n)
	\end{aligned}
\eeq
for $f$ a fixed Schwartz function (that does not depend on $\ell$ or $L$). This is the meaning of \eqref{scalingo}, and it is why we may express our main result as in Eq.~\eqref{scalingintro}, without fluid-cell averaging. 

Here and below, we keep the writing \eqref{scalingo}, which is useful for our purposes. Indeed, a simple way of establishing the distributional equation from the physical principle \eqref{meanproj} is to take the mesoscopic scale to be the ``largest possible'', $L=\ep\ell$, with the following order of limits:
\beq\label{limmeso}
	\lim_{\ep\to 0^+}\lim_{\ell\to\infty}\ell^{(n-1)d}
	\bra \overline {o_1}( z_1),\ldots, \overline {o_n}( z_n)\ket^{\rm c}_\ell.
\eeq
Smoothing out the volume $V_L$, the quantities $\overline {o_1}( z_1)$ are local observables integrated against Schwartz functions, hence the limit on $\ell$ is expected to exist. After further taking the limit $\ep\to0$, the result is $\mathsf S_{o_1,\ldots,o_n}(z_1,\ldots,z_n)$, where an explicit realisation of the distribution emerges as $\ep\to0$.

\subsubsection*{Domain of definition of Euler amplitudes}

The second comment is that we must impose restrictions on the {\em domain of definition of the Euler amplitudes}, as follows -- see Remark \ref{remashocks}.

First, for arbitrary observables $o_k$, we define the Euler amplitudes only for distinct space-time points, $(\v x_k,t_k)\neq (\v x_{k'},t_{k'})$ for all $k\neq k'$. The scaling \eqref{scalingo} still holds at colliding points, but extra distributional terms appear. By contrast, positions of conserved densities are not restricted in this way. That is, the distributional equation \eqref{scalingo} is for Schwartz functions that satisfy at least this condition:
\beq\label{Schwartzconditions}
	f(\v x_1,\ldots,\v x_n) = 0 \mbox{\em\ in a neighbourhood of }
	\{(\v x_1,\ldots,\v x_n)\in\R^{nd}:\v x_k = \v x_{k'}\}
\eeq
{\em for every $k\neq k'$ such that $t_k = t_{k'}$ and neither $o_k$ nor $o_{k'}$ is a conserved density}. This means that all observables, except conserved densities, must be at nonzero macroscopic distances and / or times from each other -- avoiding possible ``contact singularities''. 

Second, the points $(\v x_k,t_k)$ {\em must also avoid fluid singularities such as shocks}, both for generic observables, and conserved densities. We explain below why we impose these restrictions. Fluid singularities are expected to be present on a measure-zero subset of the space-time configurations $\R^{nd}$. They may be {\em intrinsic}, coming from the initial state $\bra\cdots\ket_\ell$, or, we believe, at least if $n\geq 3$, {\em extrinsic}, emerging because of the nonlinear effects observables $o_1(z_1),\ldots,o_n(z_n)$ themselves have on the state (although currently we have little understanding of the latter).

The requirement of avoiding shocks is because our derivation of the scaling fails at shocks, and further local relaxation of fluctuations also fails; and the requirement of distinct space-time positions comes from the breaking of local relaxation of fluctuations at equal space-time positions, Sec.~\ref{ssectinduced}. We show in Sec.~\ref{ssectstationary2point} an example of such contact singularity.

\subsubsection*{Derivation of Eq.~\eqref{scalingo}}

In order to establish \eqref{scalingo}, consider the following modified state, which depends on a certain number of Schwartz ``source fields'' $\lambda_k(\v x)$:
\beq\label{statestar}
	\bra \cdots\ket_\ell^{*} =
	\frc{\Big\bra \exp\Big[ \sum_{k=1}^n \int \dd^d \rx\,\lambda_k( \v\rx/\ell)
	\overline{o_k}(\v \rx/\ell,t_k)\Big] \cdots \Big\ket_\ell}{
	\Big\bra \exp\Big[ \sum_{k=1}^n \int \dd^d \rx\,\lambda_{k}( \v\rx/\ell)
	\overline{o_k}(\v\rx/\ell, t_k)\Big] \Big\ket_\ell}.
\eeq
This is still a long-wavelength state, with long-wavelength modifications of the measure inserted at macroscopic times. It is convenient to take $\lambda_k(\v x)$ to be supported on a finite neighbourhood around $\v x_k$.

By the principles of Euler hydrodynamics, for almost every space-time position $(\ell\v x,\ell t)$ with $(\v x,t)\neq (\v x_1,t_1),\,(\v x_2,t_2),\,\ldots,\,(\v x_n, t_n)$ -- in such a way that possible shocks are avoided, and local relaxation of fluctuations has occurred away from the space-time regions where the perturbations lie -- the following limit exists:
\beq\label{averagestatelambda}
	\lim_{\ell\to\infty}
	\bra \overline{o}( \v x, t)\ket_\ell^{*} =
	\mathsf o\big(\underline{\mathsf q}^*(\v x,t)\big)
\eeq
where $\underline{\mathsf q}^{*}(\v x,t)$ solves the Euler equation away from $(\v x_1,t_1),\ldots,(\v x_n,t_n)$. Of course, there is no need to use the fluid-cell averaged observable $\overline{o}$ instead of $o$ here; but this makes the derivation clearer, and does not affect the result\footnote{Speciically, on the right-hand side of \eqref{exchange}, one may put $o$ instead of $\overline{o}$, as by local space-translation invariance, one can fluid-cell average around $\ell \v x$.}.

Note that
\beq
	\int \dd^d \rx\,\lambda_k( \v\rx/\ell)
	\overline{o_k}(\v\rx/\ell, t_k)
	=
	\ell^d \int \dd^d x\,\lambda_k( \v x)
	\overline{o_k}(\v x,t_k).
%	=
%	\ell \int \dd x\,\lambda_k( x)
%	\overline{o_k}(\ell x,\ell t_k) + \mathcal o(\ell).
\eeq
Exchanging functional differentiations with respect to $\lambda_k(\cdot)$ and the limit $\ell\to\infty$, and assuming that the result is differentiable, we obtain connected correlation functions:
\beq\label{exchange}
	\frc{\delta}{\delta \lambda_1(\v x_1)}\cdots
	\frc{\delta}{\delta \lambda_n(\v x_n)}
	\mathsf o\big(\underline{\mathsf q}^{*}(\v x,t)\big)\Big|_{\lambda_1,\ldots,\lambda_n=\boldsymbol 0}
	= \lim_{\ell\to\infty} \ell^{dn} \bra \overline{o}( \v x, t), \overline {o_1}( \v x_1, t_1),\ldots ,\overline {o_n}( \v x_n, t_n)\ket^{\rm c}_\ell .
\eeq
This implies \eqref{scalingo} by re-indexing observables, and in particular it implies that it holds in a distributional sense (as we take functional derivatives).

Here we make the usual linear-response assumption, sequentially for each derivative, that although the Euler equation is in general nonlinear, it is stable (its solutions are differentiable) under linear perturbations; this is clear from the physics of mode propagations. However, the requirement of differentiability in \eqref{exchange} implies that fluid singularities must be avoided. Indeed, differentiability fails if $(\v x,t)$ lies on a singularity of the fluid emerging from the state $\bra\cdots\ket_\ell$ (such as a shock), because this singularity moves away under the insertion of source fields. Further, as source-fields insertions may introduce new such singularities, the result for $n\geq 2$ (that is, for three- or higher-point functions) may also fail for certain measure-zero configurations of space-time point. A precise analysis of conditions under which it fails is beyond the scope of this paper, hence we take the ``safe'' position that the result holds ``almost everywhere'' in space-time.

In Sec.~\ref{ssectinduced}, we explain why it is convenient to avoid colliding positions in the domain of definition of the Euler amplitudes, even though the above scaling may still be correct; see Eq.~\eqref{oS} and Remark \ref{remadistinct}. However, for conserved densities, $o_k = q_{i}$ for some $i$ and some $k$, it is possible to then extend the result to equal space-time points, by using the conservation laws. Indeed, if another observable at point $(\v x,t)$ is present within the correlation function, we may shift the time in $\overline {q_i}(\v x,t)$ as
\beqa
	\int \dd^d x\,f(\v x) (\overline {q_i}(\v x,t) - \overline {q_i}(\v x,t')) &=&
	\int \dd^d x\,f(\v x) \int_{t'}^t \dd s \,\p_s \overline {q_i}(\v x,s)\n
	&=&
	-\int \dd^d x\,f(\v x) \int_{t'}^t \dd s \,\nabla \cdot\overline {\v j_i}(\v x,s)\n
	&=&
	\int_{t'}^t \dd s \int \dd^d x\,\nabla f(\v x)\cdot \overline {\v j_i}(\v x,s)
	\label{contour}
\eeqa
thus obtaining only observables at space-time points different from $(\v x,t)$, where the scaling formula \eqref{scalingo} holds.

These arguments give the conditions as stated around Eq.~\eqref{Schwartzconditions}.

\subsubsection*{A simple example}

It is instructive to illustrate, in a very simple example, how the scaling \eqref{scalingo} indeed arises. This ballistic scaling may be surprising in diffusive systems, as one would expect diffusive scaling.

But take the example of a system with a single conserved quantity, $Q = \int \dd^d x\,q(\v x)$. For instance, this may be a generic quantum spin lattice, with $Q$ the total energy or, if spin is conserved, the total spin (in the latter case, energy and spin lie in different spin-flip sectors which decouple, and one may restrict, say, to the spin-flip odd sector, containing only the total spin). Typically, the charge density will satisfy, and large scales, a diffusive equation, and the two-point connected correlation function in a stationary state $\bra\cdots\ket_\beta$ is expected to be that of the propagation of a mode along velocity $\v 0$, with a diffusive expansion of (generically $\beta$-dependent) diffusion parameter $D$:
\beq
	\bra q (\v \rx,\micro t), q(\v 0,0)\ket_\beta^{\rm c}
	=
	\frc{\chi}{(2\pi Dt)^{d/2}}
	e^{-\v\rx^2/ (2Dt)}.
\eeq
Here $\chi = \mathsf C_{11}= \bra Q,q\ket^{\rm c}$ is the susceptibility. Then, taking the mesoscopic scale as $L = \ep\ell$ and the limit \eqref{limmeso}, we have (by translation invariance, it is sufficient to evaluate the fluid-cell mean on one observable only)
\beqa
	\lim_{\ep\to 0^+}\lim_{\ell\to\infty}\ell^{d}
	\bra \overline {q}(\v x, t), \overline {q}(\v 0,0)\ket^{\rm c}_\beta
	&=&
	\lim_{\ep\to 0^+}\lim_{\ell\to\infty}
	\frc{\ell^{d/2}\chi}{(2\pi Dt)^{d/2}}
	\frc1{|V_L|}\int_ {V_L} \dd^d\ry\,e^{-(\ell \v x + \v\ry)^2/ (2D\ell t)}\n
	&=&
	\lim_{\ep\to 0^+}\lim_{\ell\to\infty}
	\frc{\ell^{d/2}\chi}{(2\pi Dt)^{d/2}}
	\frc1{\ep^d} \int_{[-\ep/2,\ep/2]^d} \dd^d y\,e^{-\ell ( \v x + \v y)^2/ (2D t)}\n
	&=&\lim_{\ep\to 0^+}
	\frc{\chi}{\ep^d}\times \lt\{
	\ba{ll}
	1 & (\v x\in [-\ep/2,\ep/2]^d)\\
	0 & \mbox{(otherwise)}
	\ea\rt. \n
	&=& \chi\delta^d(\v x)
	\label{examplediffusive}
\eeqa
and thus we find, in this case, the distribution $S_{qq}(\v x,t;\v 0,0) = \chi\delta^d(\v x)=\chi\delta(x_1)\cdots \delta(x_d)$.

\begin{rema}\label{remader1} We emphasise that it is not possible, by arguments of functional differentiation such as those above, to get the large-scale asymptotic of correlation functions without fluid-cell averaging; or equivalently, as distributional equations on microscopic, instead of macroscopic, coordinates. Indeed, the exchange of the limit and functional derivatives in \eqref{exchange} requires that variations of source fields be on macroscopic scales.
\end{rema}

\begin{rema}\label{remader2}
Suppose in \eqref{statestar} that we use $o_k$ instead of the fluid-cell averages $\overline{o_k}$. Suppose now that we admit source fields $\lambda_k(\v{\micro x})$ that do have microscopic-scale variations. The same reasoning holds up to \eqref{exchange}. On the left-hand side we again have functional differentiations with respect to macroscopic variations only, of the source fields. But then, on the right-hand side, we would obtain not fluid-cell averaged observables, but rather averages with respect to certain microscopic-scale weight, determined by the fixed microscopic-scale shape of the source fields. This is would fall within the theory of ``structured hydrodynamics'', where the set of large-scale observables account for small-scale structures such as oscillatory behaviours. This is beyond the scope of this paper; see e.g.~\cite{buvca2019non,buca2020quantum,medenjak2020rigorous,buvca2022out,doyon2022hydrodynamic,ampelogiannis2024long}.
\end{rema}

\begin{rema} In some cases, the amplitudes are ordinary functions, such as in models of free particles, the hard rod gas, and many other integrable models in $d=1$ \cite{de2022correlation}.
\end{rema}

\subsection{Induced measure on coarse-grained densities}\label{ssectinduced}

Another crucial ingredient in the derivation of \eqref{mainintro} is the notion of induced measure on fluid-cell means of conserved densities. This is again something that was introduced in the context of the BMFT in $d=1$ \cite{doyon2023ballistic}, but is more general.

Let us construct the marginal of coarse-grained (i.e.~fluid-cell averaged) conserved densities in space-time. This marginal is induced by the initial state $\bra \cdots\ket_\ell$, along with the microscopic dynamics: transporting  the initial fluctuations by the microscopic dynamics (be it deterministic or not) gives rise to fluctuations of observables on all of space-time. Let us take the conserved densities at macroscopic coordinates. We denote the induced measure as
\beq\label{measure}
	\dbra \cdots\dket_\ell = \int [\dd \underline q]\,\re^{-\ell \mathcal M_\ell[\underline q]}\cdots.
\eeq
That is,
\beq\label{induced}
	\bra \overline {q_{i_1}}( z_1)\cdots \overline {q_{i_n}}( z_n)\ket_\ell = 
	\dbra
	q_{i_1}(z_1)\cdots q_{i_n}(z_n)\dket_\ell
\eeq
for all $\ell,L,n$ and all choices of $i_k,\,z_k=(\v x_k,t_k)$. We write the induce measure in the form $[\dd\underline q]\,\re^{-\ell \mathcal M_\ell[\underline q]}$, where formally $[\dd\underline q] = \prod_{\v x,t,i} q_i(\v x,t)$, and where the extra factor $\ell$ in the exponential $\re^{-\ell \mathcal M_\ell[\underline q]}$ is there for a reason that will become clear below. In fact, the functional $\mathcal M_\ell[\underline q]$ also depends on $L$, but we keep it implicit.

We would like to take the macroscopic limit $\ell\gg L\gg \ell_{\rm micro}$ on this measure. But what information remains in this limit?

We will argue that the induced measure \eqref{measure} leads to the following behaviour of connected correlation functions in that limit:
\beq\label{qS}
	\dbra q_{i_1}(z_1),\ldots,q_{i_n}(z_n)\dket_\ell^{\rm c}
	\sim 
	\ell^{(1-n)d}\mathsf S_{i_1,\ldots,i_n}(z_1,\cdots,z_n)
\eeq
and, away from fluid singularities and if further $z_j\neq z_{j'}$ for all $j\neq j'$,
\beq\label{oS}
	\dbra \mathsf o_1(\underline q(z_1)),\ldots,\mathsf o_n(\underline q(z_n))\dket_\ell^{\rm c}\sim
	\ell^{(1-n)d}\mathsf S_{o_1,\ldots,o_n}(z_1,\cdots,z_n).
\eeq
That is, the induced measure \eqref{measure} on coarse-grained conserved densities is such that the large-$\ell$ leading asymptotics it gives rise to for the connected correlation functions of densities equals that obtained in the original state, and that the leading asymptotics for the random variables given by the microcanonical averages of local observables $\mathsf o(\underline q)$, equals that of these observables in the original state. These are the main relations that will allow us to deduce the projection formula.

For relation \eqref{oS}, it is crucial that fluid singularities be avoided (see Remark \ref{remashocks}), and that coordinates be distinct (see Remark \ref{remadistinct} below), otherwise it is not expected to hold. The latter is the reason for imposing this constraint on the domain of the Euler amplitude, as described around Eq.~\eqref{Schwartzconditions}.

Relation \eqref{qS} is directly a consequence of \eqref{scalingo} applied to $o_k = q_{i_k}$, and the definition of the induced measure, Eq.~\eqref{induced}. But relation \eqref{oS} is non-trivial. Let us now argue for it.

The scaling \eqref{qS} is of large-deviation type, and implies that there exists $F[\underline\beta]$ such that
\beq\label{qld}
	\bdbra \exp \Big[\ell \sum_i \int \dd^{d+1} z\,\lambda^i(z)
	q_{i}( z)\Big]\bdket_\ell
	\asymp \exp \ell F[\underline \beta].
\eeq
Indeed, \eqref{qld} is equivalent to saying that $F[\underline \beta]$ is the generating functions of the Euler-scale amplitudes of conserved densities,
\beq
	F[\underline \beta] = 
	\sum_i\int \dd^{d+1} z\,\lambda^i(z) \mathsf q_i(z)+
	\frc12 \sum_{i_1,i_2}\int \dd^{d+1} z_1\int\dd^{d+1} z_2\,\lambda^{i_1}(z_1)
	\lambda^{i_2}(z_2) \mathsf S_{i_1,i_2}(z_1,z_2) +\ldots
\eeq
assuming that we can exchange functional derivatives with respect to $\lambda^i(\v x,t)$ and the limit $\ell\to\infty$ (the large-deviation principle is not broken). This further implies, via a saddle point analysis of the left-hand side of \eqref{qld} using \eqref{measure}, that the limit $\mathcal M_\infty[\underline q]$ exists\footnote{It is related to $F[\underline \lambda]$ via a Legendre transform, more precisely as
\beq
	\lambda^i(z) = \frc{\delta \mathcal M_\infty[\underline q]}{\delta q_i(z)}\Big|_{\underline q = \underline q^{[\underline\lambda]}},\quad
	F[\underline \lambda]
	= \sum_i \int \dd^{d+1} z\,\lambda^i(z)
	q_{i}^{[\underline\lambda]}(z)
	-
	\mathcal M_\infty[\underline q^{[\underline\lambda]}]
	+
	\mathcal M_\infty[\underline q^{[\underline 0]}]
\eeq
but this exact relation will not play a role here.}, that is $\ell \mathcal M_\ell[\underline q] \sim \ell \mathcal M_\infty[\underline q]$ as $\ell\to\infty$.

Similarly, from the scaling \eqref{scalingo}, there exists $G_{o_1,\ldots,o_n}[\lambda_1,\ldots,\lambda_n]$ such that
\beq
	\Big\bra \exp\Big[\ell \sum_{k=1}^n \int \dd^{d+1}z\,\lambda_k(z)
	\overline{o_k}( z)\Big]\Big\ket_\ell
	\asymp 
	\exp \ell G_{o_1,\ldots,o_n}[\lambda_1,\ldots,\lambda_n].
\eeq
Again, this is the generating functions of the Euler-scale amplitudes of the observables $o_k$. But by local relaxation of fluctuations, Eq.~\eqref{meanproj}, we may re-write the expectation value on the left-hand side as an expectation value for local observables as functions of local densities, and this, then, may be evaluated within the marginal measure \eqref{measure}. That is,
\beqa
	\Big\bra \exp\Big[\ell \sum_k \int \dd^{d+1} z\,\lambda_k(z)
	\overline{o_k}(z)\Big]\Big\ket_\ell 
	&=& \Big\bra \exp\Big[\ell \sum_k \int \dd^{d+1}z\,\lambda_k(z)
	\mathsf o_k\big(\overline{\underline q}(z)\big)+\mathcal o(\ell)\Big]\Big\ket_\ell\n
	&=& \bdbra
	\exp\Big[\ell \sum_k \int \dd^{d+1} z\,\lambda_k(z)
	\mathsf o_k\big(\underline q(z)\big)+\mathcal o(\ell)\Big]\bdket_\ell\label{lo}
\eeqa
where we have taken into consideration that the correction to local relaxation of fluctuation are vanishing as $\ell\to\infty$ (giving the $\mathcal o(\ell)$ term). Note how the $z$ integration involves a macroscopic time-averaging, thus relaxation of fluctuations can be used. On the right-hand side of the last equation of \eqref{lo}, we now have an expectation value within the measure for coarse-grained conserved densities. A saddle-point analysis, using \eqref{measure} and the existence of $\mathcal M_\infty$ established above, shows that
\beq
	\bdbra
	\exp\Big[\ell \sum_k \int \dd^{d+1} z\,\lambda_k(z)
	\mathsf o_k\big(\underline q(z)\big)+\mathcal o(\ell)\Big]\bdket_\ell
	\asymp
	\exp\ell H_{o_1,\ldots,o_n}[\lambda_1,\ldots,\lambda_n]
\eeq
where $H_{o_1,\ldots,o_n}[\lambda_1,\ldots,\lambda_n]$ is the generating function of the quantities
\[
	\lim_{\ell\to\infty} \ell^{(n-1)d}\dbra \mathsf o_1\big(\underline q(z_1)\big),\ldots,\mathsf o_n\big(\underline q(z_n)\big)\dket_\ell^{\rm c}.
\]
That is, by saddle point, the $\mathcal o(\ell)$ correction terms do not affect $H_{o_1,\ldots,o_n}[\lambda_1,\ldots,\lambda_n]$. Eq.~\eqref{lo} then implies equality
\beq
	G_{o_1,\ldots,o_n}[\lambda_1,\ldots,\lambda_n] = H_{o_1,\ldots,o_n}[\lambda_1,\ldots,\lambda_n].
\eeq
Functional differentiation gives the version of \eqref{oS} where each observable is also time-averaged. We obtain \eqref{oS} under the requirement of distinct space-time positions, as explained in Remark \ref{remadistinct}.

We will see that the combination of \eqref{qS} and \eqref{oS}, along with a generalisation of the moment-cumulant formula to partial cumulants (or partially connected correlation functions) -- Malyshev's formula -- gives rise to the general projection formula.

\begin{rema}\label{remadistinct} Local relaxation of fluctuations \eqref{meanproj} does not hold in the state \eqref{statestar} at space-time points where source fields are supported, because relaxation did not have time to occur. Thus it does not hold for ``self-nonlinear'' terms in \eqref{lo}, where a local observable, not time-averaged, is raised to a power greater than 1. We give an example of this in Sec.~\ref{ssectstationary2point}.
\end{rema}

\begin{rema} In quantum systems, fluctuations are also affected by quantum effects. However, at large scales and non-zero entropy density (non-zero temperature), one expects decoherence to occur, and the classical induced measure \eqref{measure} still to be a good description. The analysis of quantum effects via similar principles at low temperatures is an interesting question.
\end{rema}

\section{The general projection formula}\label{sectgeneralprojection}

In this section, we express our main result, the general projection formula \eqref{mainintro}, in a more precise way.

Recall that local relaxation of fluctuations \eqref{meanproj} led to the relations \eqref{qS} and \eqref{oS} that relate Euler amplitudes, defined via the asymptotics \eqref{scalingo}, to cumulants in the induced measure \eqref{measure}. Therefore, taking $q_i(z):i\in \mathfrak I,z = (\v x,t)\in \R^{d+1}$ as generating elements of the algebra $\mathcal A$ of observables for this measure, we formally are in the setup of Sec.~\ref{sectheorem}, with the replacement
\beq
	q_a,\,a\in \alpha \mbox{ (Sec.~\ref{sectheorem}) } =
	q_i(z) = q_i(\v x,t),\,i\in \mathfrak I,\,z\in\R^{d+1}\mbox{ (Sec.~\ref{sectbmft})}.
\eeq
In particular, the assumption \eqref{Soproofq} holds, with the replacement $\ell\longrightarrow \ell^d$. Therefore, the general result \eqref{generalprojectionproof} should hold.

There are a few subtleties in using \eqref{generalprojectionproof}. First, we have only defined Euler amplitudes for local observables\footnote{Although the local observables of the many-body systems form an algebra, so products are still local (Sec.~\ref{secconsqties}), this is no longer the case in the Euler scaling limit, hence for observables with respect to the measure \eqref{measure}.} $o(z)$, which, we have shown, map to functions $\mathsf o(\underline q(z))$ for cumulants in the measure \eqref{measure} when positions are distinct and away from fluid singularities. In particular, these are not generic elements of the algebra, which also includes products of such local functions at different points. Thus we only take a special case of the more general formula \eqref{generalprojectionproof}:
\beq\label{restrictiono}
	\mathsf o_k(\underline q) \mbox{ (Sec.~\ref{sectheorem}) } \mbox{ is restricted to\ \ } \mathsf o_k(\underline q(z_k))\mbox{ at distinct positions $z_k$ (Sec.~\ref{sectbmft})}.
\eeq
Second, Euler amplitudes are, in general, distributions. This makes the application of \eqref{generalprojectionproof} more subtle, see Remark \ref{remaconnection} and the discussion below. We now express the result within the present context, with these restrictions.

Recall the concept of {\em minimal connected covers} of $\{1,\ldots,n\}$, from Sec.~\ref{sectheorem}, which we denote $C^{\rm min}(n)$. Comparing with the result in the Introduction, here we think of covers of $\{1,\ldots,n\}$, instead of $\{z_1,\ldots,z_n\}$, but this is equivalent. Recall that a cover $\Upsilon = \{V_1,V_2,\ldots,V_m\}$ is a set of patches  $V_k\subseteq \{1,\ldots,n\}$ such that $\cup_k V_k = \{1,\ldots,n\}$. A connected cover is such that there is no subset of indices $K$ with $\cup_{i\in K} V_i \cap \cup_{i\not\in K}V_i = \emptyset$. So all patches are, in a sense, connected; it is possible to ``walk" between any two elements of $\{1,\ldots,n\}$, with the rule that at every step, from one element to another, the two elements must belong to a common patch. Minimality requires, however, that they overlap the least possible, and that $|V_k|\geq 2$, avoiding singletons.

Minimal connected covers are define in Eq.~\eqref{defcmax}, Sec.~\ref{sectheorem}. Here we express how to construct all minimal connected covers. This goes as follows:
\bi
\item Choose a subset $V_1$ of $\{1,\ldots,n\}$ of at least 2 elements.
\item Choose a subset $V_2$ of $\{1,\ldots,n\}$ of at least 2 elements, whose intersection with $V_1$ is a single element.
\item Choose a subset $V_3$ of $\{1,\ldots,n\}$  of at least 2 elements, whose intersection with $V_1\cup V_2$ is a single element.
\item ...
\item Choose a subset $V_m$ of $\{1,\ldots,n\}$ of at least 2 elements, whose intersection with $V_1\cup \cdots\cup V_{m-1}$ is a single element, and which exhausts $\{1,\ldots,n\}$, that is, $V_1\cup\cdots\cup V_m = \{1,\ldots,n\}$.
\ei
Then $\Upsilon = \{V_1,\ldots, V_m\}$ is a minimal connected cover. Note that $1\leq m\leq n-1$.

In formula \eqref{generalprojectionproof}, for every $V\in\Upsilon$ and $k\in V$, we construct an independent, free index variable $i^{V}_{k}\in \mathfrak I$ that runs over the set of conserved quantities of the model; so each index variable is itself indexed by the patch $V$, and the element $k\in V$ within the patch.  The result involves a summation over all free index variables. Recall that for every $k\in\{1,\ldots,n\}$, we define $\Upsilon_k\subseteq \Upsilon$ to be the set of patches that contain $k$, that is $\Upsilon_k = \{V\in\Upsilon: k\in V\}$; and that we define $m_\Upsilon(k)$ as the multiplicity of the element $k\in \{1,\ldots,n\}$ within the cover $\Upsilon$, that is $m_\Upsilon(k) = |\Upsilon_k|$ (for lightness of notation, we had denoted this $m_k$ in the Introduction). Then Eq.~\eqref{generalprojectionproof} gives, using $a = (i,z)$ and the restriction \eqref{restrictiono} (this is a more precise writing of \eqref{mainintro}):
\beq\label{generalprojection}\boxed{
\begin{aligned}
	& \mathsf S_{o_1,\ldots,o_n}(z_1,\ldots,z_n) \\
	&\quad 
	= \sum_{\Upsilon\in C^{\rm min}(n)}\ 
	\sum_{i_k^V:k\in V,V\in\Upsilon}
	\Big(\prod_{\{k_1,\ldots,k_{|V|}\}=V\in\Upsilon} \mathsf S_{i^V_{k_1},\ldots,i^V_{k_{|V|}}}(z_{k_1},\ldots,z_{k_{|V|}})\Big)\,\Big(
	\prod_{k=1}^n \frc{\p^{m_\Upsilon(k)}\mathsf o_k}{\prod_{V\in\Upsilon_k}\p \mathsf q_{i^V_k}}\Big).
	\end{aligned}
	}
\eeq
On the right-hand side, the derivatives of $\mathsf o_k=\mathsf o_k(\underline{\mathsf q})$ are to be evaluated at $\underline{\mathsf q} = \underline{\mathsf q}(z_k)$ (see \eqref{mathsfqi}), the solution to the Euler-scale equation at the macroscopic coordinate $z_k=(\v x_k,t_k)$.

For instance, we obtain
\beq\label{2ptproj}
	\mathsf S_{o_1,o_2}
	= \frc{\p \mathsf o_1}{\p \mathsf q_{i_i}}
	\frc{\p \mathsf o_2}{\p \mathsf q_{i_2}}
	\mathsf S_{i_1,i_2}
\eeq
first proposed in $d=1$ in \cite{doyon2018exact}, and
\beq\label{3ptproj}
	\mathsf S_{o_1,o_2,o_3}
	= \frc{\p \mathsf o_1}{\p \mathsf q_{i_i}}
	\frc{\p \mathsf o_2}{\p \mathsf q_{i_2}}
	\frc{\p \mathsf o_3}{\p \mathsf q_{i_3}}
	\mathsf S_{i_1,i_2,i_3}
	+
	\Bigg(\frc{\p^2 \mathsf o_1}{\p \mathsf q_{i_1}\p \mathsf q_{i_1'}}
	\frc{\p \mathsf o_2}{\p \mathsf q_{i_2}}
	\frc{\p \mathsf o_3}{\p \mathsf q_{i_3}}
	\mathsf S_{i_1,i_2}
	\mathsf S_{i_1',i_3}
	+
	\mbox{cyclic permutations $1\to 2\to 3$}
	\Bigg)
\eeq
first obtained, again in $d=1$, in \cite{doyon2023ballistic}. Here, the implicit space-time positions are $z_1$, $z_2$ and $z_3$ for indices $i_1$, $i_2$ and $i_3$, respectively.

Eq.~\eqref{generalprojection} is a formula for the Euler-scale amplitudes. On the left-hand side, multiplying by $\ell^{(1-n)d}$, we obtain the leading asymptotic of the physical correlation function (in a distributional sense). On the right-hand side, we must multiply by $\prod_{V\in\Upsilon}\ell^{(1-|V|)d} = \ell^{(|\Upsilon|-\sum_{V\in\Upsilon}|V|)d}$. We show in Subsection \ref{ssectproj} that for the set of minimal connected covers, $|\Upsilon|-\sum_{V\in\Upsilon}|V|=1-n$ (it is relatively clear from the construction method above). Thus we have $\ell^{(1-n)d}$ for each term, and therefore the same projection formula \eqref{generalprojection} holds for the physical correlation functions themselves, instead of the amplitudes, with equality replaced by equality of leading asymptotics $\sim$ as $\ell\to\infty$.

As mentioned in the introduction, on the right-hand side of \eqref{generalprojection}, we have a product of distributions. Physically, one expects the distributional part to only occur when position, say, $\v x_k$, ballistically transported to time $t_k$, collide with position $\v x_{k'}$ transported to time $t_{k'}$. Thus, this involves two points; distributional parts involving a single point $z_k$ would correspond to fixed singularities (such as impurities) in space-time. As every two patches in a minimal connected cover can only share at most one point, this does not lead to any problem, and the product makes sense. We will see in examples below how this works. However, we do not have general a general proof yet of this.

An important case, where the formula simplifies slightly, is that where one observable $o$ is arbitrary, while the others are conserved densities:
\beq
	\mathsf S_{o,q_{i_1},\ldots,q_{i_{n-1}}}(z,z_1,\ldots,z_{n-1}).
\eeq
In this case, thinking in terms of covers of the positions $z,z_1,\ldots,z_{n-1}$, only $z$ may have multiplicity greater than 1. Thus the set of minimal connected covers is obtained by choosing a partition $\Gamma$ of $\{1,\ldots,n-1\}$, and forming the cover with patches $V = \{z_k:k\in W\}\cup \{z\}$ for all $W\in\Gamma$. In this way, there is a bijective correspondence between the minimal connected covers involved, and partitions $P(n-1)$ of $\{1,\ldots,n-1\}$. That is, we obtain
\beqa\label{specialprojection}
	\lefteqn{\mathsf S_{o,q_{i_1},\ldots,q_{i_{n-1}}}(z,z_1,\ldots,z_{n-1})}\\
	&& =
	\sum_{\Gamma\in P(n-1)}
	\sum_{i^W:W\in \Gamma}
	\Big(\prod_{\{k_1,\ldots,k_{|W|}\}=W\in \Gamma}
	\mathsf S_{i^W,i_{k_1},\ldots,i_{k_{|W|}}}(z,z_{k_1},\ldots,z_{k_{|W|}})\Big)\,\Big(
	\frc{\p^{|\Gamma|}\mathsf o}
	{\prod_{W\in\Gamma}\p \mathsf q_{i^W}}\Big)\no
\eeqa
where the free indices $i^W$ for the partition $\Gamma$, run over all conserved quantities.

\begin{rema}\label{remaconnection} Sec.~\ref{sectheorem}, and App.~\ref{appproof}, give the precise context in which \eqref{generalprojection} holds, and its proof. The proof is based on the scaling relations \eqref{qS}, and show that \eqref{oS} in fact follows. In particular, it is important that the first is to hold also at equal points; we argued above that this is the case. However, there are two limitations:
\bi
\item[(1)] Relations \eqref{qS} and \eqref{oS} are distributional, while the proof assumes that they are true asymptotic relations. Thus, in order to apply the proof, one should consider appropriate functional statements, where, e.g., $q_{i_k}(z_k)$ in \eqref{qS} are replaced by $\ep^{-d} \int \dd^d x\,w(\v x/\ep) q_{i_k}(z_k)$ for some unit-normalised Schwartz weight function $w(\v y)$ supported on $\v y\in[-1/2,1/2]$. For $\ep>0$ small enough, relaxation of fluctuations \eqref{meanproj} holds, with small correction in $\ep$. Thus the result \eqref{generalprojection}, under this replacement, holds, with small corrections in $\ep$. If the right-hand side of \eqref{generalprojection} makes sense as a distribution, even though it is a product of distributions, then the limit $\ep\to0$ indeed is expected to give this right-hand side. If this is not the case, small-$\ep$ corrections may combine with diverging terms as $\ep\to0$ (corresponding to un-defined products of distribution), invalidating the result. Thus, with these arguments, we show \eqref{generalprojection} under the caveat that the right-hand side makes sense as a distribution. We have argued above that this must be the case in general.
\item[(2)] The proof applies to the set of observables such that $\mathsf o(\mathsf q)$ are {\em polynomials} in the conserved densities. This is in order to avoid discussions of convergence. As the result involves only a finite number of derivatives, one may expect convergence not to be a problem; but a full analysis is beyond the scope of this paper.
\ei
\end{rema}

\section{Euler amplitudes in space-time stationary states}\label{sectapplications}

We now establish the explicit form of the Euler amplitudes in stationary states. By our projection formula, it is sufficient to concentrate on conserved densities, $S_{i_1,\ldots,i_n}(z_1,\ldots,z_n)$; recall that $z_k = (\v x_k,t)$ are space-time points in macroscopic coordinates. This is done by solving the dynamical equations obtained from the conservation laws. Such equations become closed equations for the Euler amplitudes of conserved densities, by the application of the projection formula \eqref{specialprojection} for current observables $o=j_{ir}$ (with $r=1,\ldots,d$). For instance, one has
\beq\label{evolamplitude}
	\p_{t_1} \mathsf S_{i_1,\ldots,i_n}(z_1,\ldots,z_n)
	+ \nabla_1\cdot \mathsf S_{\v j_{i_1},q_{i_2},\ldots,q_{i_n}}(z_1,\ldots,z_n)
	= 0
\eeq
where $\nabla_1 =  \p_{\v x_1} = (\p_{x_{11}},\ldots,\p_{x_{1d}})$, and then one may use \eqref{specialprojection} for $\mathsf S_{\v j_{i_1},q_{i_2},\ldots,q_{i_n}}(z_1,\ldots,z_n)$ -- this is valid in general, not just in stationary states.

The results established below for three-point functions hold under the caveat that there are no {\em extrinsic} fluid singularities everywhere in space-time  -- see the discussion below Eq.~\eqref{Schwartzconditions} in Sec.~\ref{ssecteulerscaling}. Indeed, we integrate projection results in space-time to obtain the formula. This is guaranteed at least in systems where no shocks emerge, such as linearly degenerate systems \cite{rozlinedeg,liu1979development,bressan2013hyperbolic,doyon2025hydrodynamic}.

\subsection{Two-point amplitudes}\label{ssectstationary2point}

In space-time stationary states
\beq
	\bra\cdot\ket = \bra\cdot\ket_{\underline\beta},
\eeq
the structure of 2-point amplitudes of conserved densities, and some aspects of amplitudes for general observables, is well studied. For completeness and as this is instructive, we develop it here from our formalism.

Using \eqref{2ptproj} and \eqref{Amatrix}, we obtain for 2-point amplitudes:
\beq
	\p_t \mathsf S_{ij}(\v x,t;\v 0,0) +
	\sum_k\v{\mathsf A}_i^{~k} \cdot\nabla \mathsf S_{kj}(\v x,t;\v 0,0)= 0.
\eeq
The initial condition at $t=0$ can be obtained by a calculation similar to \eqref{examplediffusive}, and is given by $\mathsf S_{ij}(\v x,0;\v 0,0) = \delta^{d}(\v x)\mathsf C_{ij}$. This makes sense as the domain of Euler amplitudes for conserved densities includes equal space-time positions. Then, the above is solved as
\beq\label{2ptconservedstat}\boxed{
	\mathsf S_{ij}(\v x,t;\v 0,0) =
	\sum_l
	\int \frc{\dd^d p}{2\pi}\,
	\Big(\exp {\ri \v p\cdot (\v x-\v {\mathsf A}t)}\Big)_{i}^{~l}\,
	\mathsf C_{lj}.}
\eeq
For generic observables, using the projection formula \eqref{2ptproj} and \eqref{deroq}, we obtain
\beq\label{2ptamplitudestat}
	\mathsf S_{o_1,o_2}(\v x,t;\v 0,0) =
	\sum_{ijl} \bra Q_i,o_1\ket^{\rm c}\,
	\mathsf C^{ij}
	\int \frc{\dd^d p}{2\pi}\,
	\Big(\exp {\ri \v p\cdot (\v x-\v {\mathsf A}t)}\Big)_{j}^{~l}\,
	\mathsf \bra Q_l,o_2\ket^{\rm c}.
\eeq
We believe this is the first time this formula is established in $d>1$, although it has the same structure as in $d=1$. Note that in $d=1$, we have 
\beq\label{2ptspecialisation}
	\int \frc{\dd^d p}{2\pi}\,
	\exp {\ri \v p\cdot (\v x-\v {\mathsf A}t)}
	\stackrel{d=1}= \delta(x - \mathsf A t)
\eeq
while in $d\geq 1$ with a single conserved quantity, writing $\v{\mathsf A} = \v{\mathsf v}$ for the velocity vector (the vector of 1 by 1 matrices $\mathsf A_\mu = \mathsf v_\mu$, $\mu=1,\ldots,d$), we have
\beq\label{2ptspecialisationsd}
	\int \frc{\dd^d p}{2\pi}\,
	\exp {\ri \v p\cdot (\v x-\v {\mathsf A}t)}
	\stackrel{\text{one\ conserved\ quantity}}= \delta^{d}(\v x - \v{\mathsf v} t).
\eeq
In general, recalling Eq.~\eqref{Ap} and the discussion around it,
\beqa
	\int \frc{\dd^d p}{2\pi}\,
	\exp {\ri \v p\cdot (\v x-\v {\mathsf A}t)}
	&=&
	\int \frc{\dd^d p}{2\pi}\,
	\re^{\ri |\v p| \,(\h{\v p}\cdot \v x- \mathsf A(\h{\v p})t)}\n
	&=&
	\int_{\mathbb S^{d-1}} \!\!\!\!\dd\Omega(\h{\v p})\,
	\int_0^{\infty} \frc{\dd p\,p^{d-1}}{2\pi}\,
	\re^{\ri p \,(\h{\v p}\cdot \v x- \mathsf A(\h{\v p})t)}\n
	&=&
	\int_{\mathbb S^{d-1}_+} \!\!\!\!\dd\Omega(\h{\v p})\,
	\int_{-\infty}^{\infty} \frc{\dd p\,|p|^{d-1}}{2\pi}\,
	\re^{\ri p \,(\h{\v p}\cdot \v x- \mathsf A(\h{\v p})t)}\n
	&=&
	\lt\{\ba{ll}
	\int_{\mathbb S^{d-1}_+} \!\!\!\!\dd\Omega(\h{\v p})\,
	(-\ri)^{d-1}\delta^{(d-1)}(\h{\v p}\cdot \v x- \mathsf A(\h{\v p})t)
	& \mbox{($d$ is odd)}\z
	\int_{\mathbb S^{d-1}_+} \!\!\!\!\dd\Omega(\h{\v p})\,
	\frc{i^d(d-1)!}{\pi}\,\prin\,\frc{1}{(\h{\v p}\cdot \v x- \mathsf A(\h{\v p})t)^{d}}
	& \mbox{($d$ is even)}
	\ea\rt.
\eeqa
where $\mathbb S^{d-1}$ is the $(d-1)$-dimensional unit sphere, $\dd\Omega(\h {\v p})$ is the surface element on it, and $\mathbb S^{d-1}_+$ is the right half $p_1>0$ of the $(d-1)$-dimensional unit sphere. Here $\prin$ indicates Cauchy principal value, and $\delta^{(n)}(x) = \dd^n \delta(x)/\dd x^n$. Using normal modes \eqref{normalmodes} in this expression, we see that $\{v_I(\h{\v p})\}_I$ indeed can be interpreted as the set of hydrodynamic velocities in the direction $\h{\v p}$.

\subsubsection*{Contact singularities}

For generic observables $o_1,o_2$, not conserved densities, formally specialising \eqref{2ptamplitudestat} to $t=0$ {\em does not give the distribution corresponding to the ballistic limit of the equal-time 2-point correlation function} -- this is an example of a contact singularity, and is why we restricted the domain of the amplitudes for generic observables as is explained around \eqref{Schwartzconditions}. Indeed, a direct calculation similar \eqref{examplediffusive} gives 
\beq
	\lim_{\ell\to\infty}
	\ell \bra \overline{o_1}(\v x,0)\overline{o_2}(0,0)\ket^{\rm c}
	= \bra O_1,o_2\ket^{\rm c}\, \delta^d(\v x)
\eeq
where
\beq
	\bra O_1,o_2\ket^{\rm c}
	=
	\int \dd^d \rx\,\bra o_1(\v x),o_2(\v 0)\ket^{\rm c}
	\neq
	\sum_{ij} \bra Q_i,o_1\ket^{\rm c}\,
	\mathsf C^{ij}
	\mathsf \bra Q_j,o_2\ket^{\rm c}.
\eeq
The discrepancy arises because relaxation of fluctuations, Eq.~\eqref{meanproj}, does not hold at colliding space-time points (see Remark \ref{remadistinct}). At equal times and positions, relaxation has not had time to occur independently on the two coarse-grained observables, so it cannot be used.

\subsubsection*{Drude weight}

At different times, relaxation has occurred,
\beq
	\lim_{\ep\to0}\lim_{\ell\to\infty}
	\ell \bra \overline{o_1}(\v x,\ep),\overline{o_2}(0,0)\ket^{\rm c}
	= 
	\sum_{ij} \bra Q_i,o_1\ket^{\rm c}\,
	\mathsf C^{ij}
	\mathsf \bra Q_j,o_2\ket^{\rm c}\,\delta^d(\v x).
\eeq
This is a manifestation of a projection formula shown rigorously in \cite{doyon2022hydrodynamic,ampelogiannis2024long}:
\beqa
	\lim_{\micro t\to\infty}
	\int \dd^d \rx\,\bra o_1(\v{\rx},\micro t),o_2(\v 0,0)\ket^{\rm c} &=&
	\lim_{\ell\to\infty}
	\int \dd^d \rx\,\bra o_1(\v{\rx},\ell \ep),o_2(\v 0,0)\ket^{\rm c} \n &=&
	\lim_{\ell\to\infty}
	\ell \int \dd^d x\,\bra \overline{o_1}(\v x,\ep),\overline{o_2}(\v 0,0)\ket^{\rm c}\n
	 &=&
	 \sum_{ij} \bra Q_i,o_1\ket^{\rm c}\,
	\mathsf C^{ij}
	\mathsf \bra Q_j,o_2\ket^{\rm c}
\eeqa
where we may choose any $\ep\neq0$. For $o_1$ and $o_2$ being current observables $j_{i_1}$ and $j_{i_2}$, this gives an exact projection formlula for the {\em Drude weight}.

\subsection{Three-point amplitudes}\label{ssectstationary3point}

Three-point amplitudes can be obtained  from \eqref{3ptproj} in a similar fashion. The result is rather complicated in arbitrary dimension and with arbitrary number of conserved quantities. Here, we express it in $d=1$ for arbitrary set of conserved quantities, and in arbitrary $d$, for a single conserved quantity.

\subsubsection*{$d=1$, arbitrary number of conserved quantities}

In $d=1$ with one or more conserved quantities, the transformation to normal modes is given by Eq.~\eqref{normalmodes1d} (see App.~\ref{appmatrices}). Recall that we use capital letters $I,J,K,\ldots$ for the normal mode basis, and that the transformation matrix $\mathsf R_I^{~i}$ depends on the stationary state considered. We are interested in the three-point amplitude
\beq
	\mathsf S_{IJK}(z_1,z_2,z_3) = \sum_{ijk}\mathsf R_I^{~i}\mathsf R_J^{~j}\mathsf R_K^{~k}\,\mathsf S_{ijk}(z_1,z_2,z_3),\quad z_n = (x_n,t_n).
\eeq

Using \eqref{3ptproj}, \eqref{evolamplitude}, \eqref{2ptconservedstat} and \eqref{2ptspecialisation}, we have to solve (no summation over indices)
\beq\label{eqSIJK}
	(\p_{t_1}+ \mathsf v_I \p_{x_1}) \mathsf S_{IJK} 
	+ \mathsf A_{I}^{~JK} \p_{x_1}\big(\delta(x_{12}-\mathsf v_J t_{12}) \delta(x_{13}-\mathsf v_K t_{13})\big) = 0
\eeq
where $x_{ij} = x_i-x_j$, $t_{ij} = t_i-t_j$ and the 3-point coupling is, see \eqref{ominus} and \eqref{Atensornormal1d},
\beq
	\mathsf A_{I}^{~JK} = \bra Q_J,Q_K,j_I^-\ket^{\rm c}.
\eeq
We simultaneously have to solve two other equations obtained by the above under the cyclic permutation $I\to J\to K$, $x_1\to x_2\to x_3$ and $t_1\to t_2\to t_3$. The initial condition is given by
\beq\label{initSIJK}
	\mathsf S_{IJK}(x_1,0;x_2,0;x_3,0) = \delta(x_{12})\delta(x_{13})\bra Q_I,Q_J,q_K\ket^{\rm c}
\eeq
where
\beq
	\bra Q_I,Q_J,q_K\ket^{\rm c}=\sum_{ijk} \mathsf R_I^{~i}\mathsf R_J^{~j}\mathsf R_K^{~k}\int \dd \rx\dd\ry\, \bra q_i(\rx),q_j(\ry),q_k(0)\ket^{\rm c}.
\eeq

The system of equations \eqref{eqSIJK} (and its cyclic permutations) along with the initial condition \eqref{initSIJK} is expected to have a unique solution. We show in App.~\ref{app3pt} that the following is a solution:
\beq\label{solSIJK}\boxed{
	\mathsf S_{IJK} = \bra Q_I,Q_J,q_K\ket^{\rm c}\delta(u_{12})\delta(u_{13})
	-
	\big(\mathsf A_{I}^{~JK}
	B_{t_1}(u_{12},u_{13},\mathsf v_{IJ},\mathsf v_{IK}) + \mbox{cyclic permutations}\big)}
\eeq
where $u_a = x_a -\mathsf v_a t_a$ with $\mathsf v_1 = \mathsf v_I,\,\mathsf v_2 = \mathsf v_J,\, \mathsf v_3=\mathsf v_K$, and $u_{ab} = u_a-u_b$, $\mathsf v_{ab} = \mathsf v_a - \mathsf v_b$, and where $B_t(u,u',v,v') = b_t(u,u',v,v') - b_0(u,u',v,v')$ and
\beq\label{solSIJK2}
	b_t(u,u',v,v') =\lt\{\ba{ll}
	\delta'(v'u-vu')\Big(|v'|s(u'+v't) - |v|s(u+vt)\Big) & (v\neq 0,\,v'\neq 0)\z
	\frc1{v'}\Big(\delta(u'+v't)\delta(u) + s(u'+v't)\delta'(u)\Big)
	& (v=0,\,v'\neq 0)\z
	\frc1{v}\Big(\delta(u+vt)\delta(u') + s(u+vt)\delta'(u')\Big)
	& (v\neq 0,\,v'= 0)\z
	t \Big(\delta'(u)\delta(u') + \delta(u)\delta'(u')\Big) & (v=v'=0)
	\ea\rt.
\eeq
where $s(x) = \frc12 \sgn(x)$ and $\delta'(x) = \dd \delta(x)/\dd x$.

Because the state $\bra\cdots\ket$ is space-time stationary, the three-point amplitude should be invariant,
\beq
	\mathsf S_{IJK}(x_1+y,t_1+r;x_2+y,t_2+r;x_3+y,t_3+r)
	=
	\mathsf S_{IJK}(x_1,t_1;x_2,t_2;x_3,t_3),
\eeq
under space translations $y$ and time translations $r$. The solution \eqref{solSIJK} is manifestly space-translation invariant, as it only depends on differences of positions. However, it is not manifestly time-translation invariant. As explained in App.~\ref{app3pt},
the fact that it indeed has this invariance follows from the fundamental property \eqref{sym3pt1d} of the 3-point coupling $\mathsf A_I^{~JK}$. This arises thanks to the very special combination of terms in \eqref{solSIJK}, including the particular form of the initial condition. This therefore confirms the internal self-consistency of our general projection formalism.

In the case of a single conserved quantity $Q = \int \dd\rx\, q(\rx)$, with $q$ and $j$ the conserved density and current, there is no need to go to normal coordinates. We denote the susceptibility as $\chi$, that is
\beq
	\chi = \mathsf C_{11} = \bra Q,q\ket^{\rm c},\quad
	\chi' = \frc{\p\chi}{\p\mathsf q} =  \chi^{-1} \bra Q,Q,q\ket^{\rm c}
\eeq
and $\mathsf A = \p \mathsf j/\p\mathsf q = \mathsf v$ is the single fluid velocity, with $\mathsf v' = \p \mathsf v / \p \mathsf q$. The solution simplifies to
\beq\label{S31d1q}
	\mathsf S_3 = \chi\chi'\delta(u_{12})\delta(u_{13})
	- \chi^2 \mathsf v' \Big(t_1 \p_{x_1} \big(\delta(u_{12})\delta(u_{13})\big)
	+ \mbox{cyclic permutations}\Big)
\eeq
where here and below, for a single conserved quantity, we use the notation $\mathsf S_n = \mathsf S_{\underbrace{q\cdots q}_{n\,\text{times}}}$.

\subsubsection*{$d>1$, one conserved quantity}

In higher dimensions $d>1$, we restrict to a single conserved quantity only. Again there is no need to go to normal coordinates. We have the evolution equation
\beq
	\p_{t_1} \mathsf S_3 + \v {\mathsf v}\cdot \nabla_1 \mathsf S_3
	+\v {\mathsf v}'\cdot\nabla_1 \big(
	\delta^{d}(\v x_{12} - \v{v}t_{12})
	\delta^{d}(\v x_{13} - \v{v}t_{13})\chi^2
	\big)=0
\eeq
whose solution is the higher-dimensional version of \eqref{S31d1q},
\beq
	\mathsf S_3 = \chi\chi'\delta^{d}(\v{u}_{12})\delta^{d}(\v{u}_{13})
	- \chi^2\v{\mathsf v}' \cdot \Big(t_1 \nabla_{1} \big(\delta^{d}(\v{u}_{12})\delta^{d}(\v{u}_{13})\big)
	+ \mbox{cyclic permutations}\Big)
\eeq
where we recall that
\beq
	\chi = \bra Q,q\ket^{\rm c} = \int \dd^d \rx \,
	\bra q(\v\rx),q(\v 0)\ket^{\rm c},\quad
	\chi\chi' = \bra Q,Q,q\ket^{\rm c}
	= \int \dd^d \rx \dd^d \ry\,
	\bra q(\v\rx),q(\v\ry),q(\v 0)\ket^{\rm c}
\eeq
and  $\v{u}_{ij} = \v x_{ij} - \v{\mathsf v} t_{ij}$. The fact that this solution is time-translation invariant follows from the same argument as that for the case $v=v'$ in App.~\ref{app3pttime}.

\subsection{Example: TASEP}\label{ssectexample}

For illustration, we consider a simple example where there is hope that correlation functions may be evaluated numerically or analytically: the totally asymmetric simple exclusion process (TASEP). See for instance \cite{liggett1985interacting}. As mentioned above, these results hold under the caveat that there should be no {\em extrinsic} fluid singularities, Sec.~\ref{ssecteulerscaling}. As we currently do not have a good understanding of such possible singularities, and the TASEP hydrodynamic is not linearly degenerate, Eq.~\eqref{resTASEP} below is conjectural; it would be very interesting to verify it or to understand its corrections, if any.

TASEP is a stochastic model for the dynamics of non-overlapping particles hopping on $\Z$: each site $\rx\in\Z$ is either occupied -- contains a particle -- or not. This model admits a single conserved quantity: the total number of particles (occupied sites). We denote the corresponding density by
\beq
	q(\rx,\micro t) = 1\ \mbox{($\rx$ is occupied at time $\micro t$)},\ 
	0\ \mbox{(otherwise).}
\eeq
The stationary states are the product measures, with each site having occupation probability $\rho\in[0,1]$. The density of occupation per unit site in a stationary state is therefore simply
\beq
	\mathsf q = \bra q(\rx,\micro t)\ket = \rho.
\eeq
Thermodynamic quantities are known exactly, with the susceptibility and average current given by the same expression:
\beq
	\chi = \mathsf j = \rho(1-\rho).
\eeq
The velocity and derivative of the susceptibility are $\chi'=\mathsf v = 1-2\rho$, and $\mathsf v' = -2$. Putting these expressions in \eqref{S31d1q}, we get
\beqa
	\lefteqn{\bra q(\ell x_1,\ell t_1),q(\ell x_2,\ell t_2),q(\ell x_3,\ell t_3)\ket^{\rm c}} &&\n
	&\sim& \ell^{-2}\Big[ 
	\rho(1-\rho)(1-2\rho)\delta(x_{12} - \mathsf vt_{12})\delta(x_{13} - \mathsf vt_{13}) + \n
	&& \qquad +\, 2\rho^2(1-\rho)^2 \Big(t_1 \p_{x_1} \big(\delta(x_{12} - \mathsf vt_{12})\delta(x_{13} - \mathsf vt_{13})\big)
	+ \mbox{cyclic permutations}\Big)\Big].\label{resTASEP}
\eeqa

\subsection{Example: many-body integrable systems}\label{ssectexample2}

The hydrodynamics of many-body integrable systems is generalised hydrodynamics (GHD) \cite{doyon2020lecture,bastianello2022introduction,essler2023short,spohn2024hydrodynamic}. It is known at least in many cases not to develop shocks or other fluid singularities \cite{doyon2017large,hubner2024new,hubner2024existence}, hence for which our general results for three-point functions are more solidly established.

We refer to \cite{doyon2020lecture} for the ideas and formalism of GHD. In this context, there are in fact three natural bases: the ``ordinary'' basis of conserved quantities $Q_i$ for $i=1,2,3,\ldots$, such that a single asymptotic particle of spectral parameter $\lambda\in\mathcal S$ carries a quantity $h_i(\lambda)$ (for some set of functions $h_i(\cdot)$); the basis $Q_\theta$ for $\theta\in\mathcal S$, where the quantity carried is $\delta(\theta-\lambda)$; and the normal-mode basis $Q_I$ again for $I\in\mathcal S$. In many models the spectral space $\mathcal S$ is simply $\mathcal S=\R$, while in some models it is $\R^2$, in others $\R^+$, etc. In particular, in \cite{doyon2020lecture}, formula (3.75) gives the covariance matrix $\mathsf C_{ij}$ and (3.84) the covariance matrix $\mathsf C_{\theta,\theta'}$, formula (3.82) the flux Jacobian $\mathsf A_\theta^{~\theta'}$, and formula (3.86) the {\em un-normalised} matrix $\mathsf R_I^{~\theta}$ of transformation to linear normal modes (that is, it is not normalised to give $\mathsf C_{IJ} = \delta_{IJ}$). The normalised one is
\beq
	\mathsf R_I^{~\theta} = \frc1{\sqrt{\rho_If_I}}(1-nT)_{I,\theta}
\eeq
where $T$ is the scattering kernel, $\rho$ the quasi-particle density, $f$ the statistical factor, and $n$ the occupation function (a Riemann invariant) -- see \cite{doyon2020lecture}.

The calculations of $\bra Q_I,Q_J,q_K\ket^{\rm c}$ and $\mathsf A_{I}^{~JK}$ are done by applying derivatives on the $\mathsf C$ and $\mathsf A$ matrices, according to their basic definitions. For this purpose, formulae (3.73) and (3.93) in \cite{doyon2020lecture} are useful. In normal modes, one finds
\beq
	\mathsf A_I^{~JK} = T_{IJ}^{\rm dr}(v^{\rm eff}_J-v^{\rm eff}_I)
	\frc{n_I\sqrt{\rho_Jf_J}}{\rho_I}\delta(I-K) + \big(J\leftrightarrow K\big)
\eeq
where $T_{IJ}^{\rm dr}$ is the dressed scattering kernel \cite[Eq 3.46]{doyon2020lecture} and $v^{\rm eff}_I$ the effective velocity \cite[Eqs 3.61, 3.63]{doyon2020lecture}. An expression for $\bra Q_I,Q_J,q_K\ket^{\rm c}$ is also found, and putting everything together back to the ordinary basis, we obtain
\beqa\label{solSijkinteg}
	\mathsf S_{ijk} &=& \int \dd\theta\dd\theta'\dd\theta''\,h_i^{\rm dr}(\theta)h_j^{\rm dr}(\theta')h_k^{\rm dr}(\theta'')\times\Big( \n
	&&
	\quad \big(\ f_\theta(1-nT)^{-1}_{\theta,\theta''}
	\rho_{\theta''}f_{\theta''}
	\delta(\theta-\theta')+\text{cyclic permutations}\ \theta\to\theta'\to\theta''\n &&
	\quad \quad +\;\rho_\theta f_\theta n_\theta (f'_\theta-2f_\theta)\delta(\theta-\theta')
	\delta(\theta-\theta'')
	\ \big)\;
	\delta(u_{12})\delta(u_{13})\n &&
	\quad -\;
	\big(\ 
	\big[\ 
	T^{\rm dr}_{\theta,\theta'}(v^{\rm eff}_{\theta'}-v^{\rm eff}_{\theta})
	\rho_{\theta'}f_{\theta'}n_\theta f_\theta \delta(\theta-\theta'')
	B_{t_1}(u_{12},u_{13},v^{\rm eff}_\theta-v^{\rm eff}_{\theta'},0)
	\n && \qquad\qquad +\;
	T^{\rm dr}_{\theta,\theta''}(v^{\rm eff}_{\theta''}-v^{\rm eff}_{\theta})
	\rho_{\theta''}f_{\theta''}n_\theta f_\theta \delta(\theta-\theta')
	B_{t_1}(u_{12},u_{13},0,v^{\rm eff}_\theta-v^{\rm eff}_{\theta''})
	\ \big]
	\n && \qquad\quad +\;
	\mbox{cyclic permutations}\ \big)
	\quad \Big)
\eeqa
where $f'_\theta = \dd f_\theta/\dd\theta$ and ${}^{\rm dr}$ is the dressing operation \cite[Eq 3.33]{doyon2020lecture}. Note how it is only the 2nd and 3rd case of \eqref{solSIJK2} that are involved, because of the delta-functions involved in $\mathsf A_I^{~JK}$.

\section{Conclusion}\label{sectconclusion}

In this paper, we have derived a nonlinear projection formula, relating dynamical correlation functions of local observables in finite-entropy states, to those of the conserved densities admitted by the many-body system. The derivation is based on the principle of local relaxation of fluctuations, an implementation of the projection ideas of Mori and Zwanzig \cite{mori1965transport,zwanzig1966statistical} and a nonlinear version of the Boltzmann-gibbs principle, introduced in the context of the ballistic macroscopic fluctuation theory (BMFT) \cite{doyon2023emergence,doyon2023ballistic}. It holds for $n$-point correlation functions in $d$ dimensions of space, for all $n\geq 1$ and $d\geq 1$, and, we expect, for any local dynamics, such as Hamiltonian, circuit or stochastic, both quantum and classical. The formula is expressed as a sum over minimal connected covers, making it easy to work out graphically for any specific $n$. This generalises to arbitrary $n$ and $d$ various special cases established earlier, including the well-known projection formula for two-point functions, and makes it clear how nonlinear effects arise in higher-point functions.

Using conservation laws and the projection formula for three-point functions involving currents, we have established the form of three-point correlation functions of conserved densities (and therefore, by projection, of any local observables) in various cases, solely in terms of the thermodynamic and Euler hydrodynamic data: the flux Jacobian and covariance matrix. We have shown that the result is space- and time-translation invariant, the latter requiring special cancellations of terms along with a new symmetry relation for ``3-point coupling'' that we have established Eq.~\eqref{sym3pt}. It would be interesting to analyse higher-point functions in stationary states, including cumulants of total currents through surfaces in arbitrary dimensions. It would be particularly interesting to study long-range correlations emerging in non-stationary states, by techniques similar to those of \cite{doyon2023emergence}. A natural question is to extend the present formalism to higher-dimensional integrable systems, following the soliton-gas picture where fundamental ``particles'' are extended objects \cite{bonnemain2025two}. The structure of a sum over minimal connected covers is a kind of generalisation of Wick's theorem for Gaussian fields. It suggests that Euler-scale correlation functions may be re-organised into a non-Gaussian field-theoretic formulation. The BMFT does this for linearly degenerate systems, including integrable systems, however in general shocks may arise, which have to be taken into account. In particular, on trajectories in space-time where shocks lie, the projection formula is not expected to hold, because the principle of relaxation of fluctuations breaks down. It would be interesting to work out the consequences of these on explicit expression of correlation functions.

Perhaps the most interesting next steps are to extend the present formalism beyond the ballistic (Euler) scale (see the companion paper \cite{doyon2025hydrodynamic}), as is done in \cite{PhysRevLett.134.187101} for integrable systems, and to include quantum fluctuations, see e.g.~\cite{ruggiero2020quantum,urilyon2025quantum}.

On this aspect, we note that in $d=1$ integrable systems, for $r=1$, $r=2$ , $r=3$, etc, contributions at order $\ell^{1-n-k}$ to $n$-point correlation functions that satisfy
\beq
	k+n \leq r,\quad n\geq 1,\ k\geq 0
\eeq
are related to each other in the way that is suggested in \cite{PhysRevLett.134.187101}:
\bc
\begin{tabular}{c|cccc}
$n$\ \textbackslash\  $k$ & 0th order (ballistic) & 1st order (diffusive) & 2nd order (dispersive) & $\cdots$ \\
\hline
1-point function&$\star$&$\circ$&$\bullet$&$\cdots$\\
2-point function&$\circ$&$\bullet$&&\\
3-point function&$\bullet$&&&\\
$\vdots$&$\vdots$&&&\\
\end{tabular}
\ec
In particular, the leading behaviour of $n$-point functions affect the $n$th order of the hydrodynamic expansion of one-point functions. More generally, in generic $d$ dimensional systems, we expect contributions at order $\ell^{(1-n)d-k}$ to $n$-point correlation functions that satisfy
\beq
	k+nd \leq rd,\quad n\geq 1,\ k\geq 0
\eeq
are related to each other, but in general with additional stochastic contributions (as explained in \cite{doyon2025hydrodynamic} for linearly degenerate systems).

\medskip

{\bf Acknowledgments.}
I am grateful to Jacopo De Nardis, Friedrich H\"ubner and Takato Yoshimura for discussions and works related to the results presented here. This work was supported by EPSRC under grants EP/W010194/1, and EP/Z534304/1 (ERC advanced grant scheme).

\appendix

\section{Hydrodynamic matrices and tensors}\label{appmatrices}

In this appendix, we exand on the standard hydrodynamic matrices used in describing the Euler scale in arbitrary dimension. We also prove a relation, which is essential in order to understand three-point amplitudes. Recall Sec.~\ref{secconsqties}.

A useful quantity is the {\em flux Jacobian vector} $\v{\mathsf A} = (\mathsf A_1,\ldots,\mathsf A_d)$, the vector of matrices obtained by differentiating the fluxes (average currents):
\beq\label{Amatrix}
	\v{\mathsf A}_i^{~l}:=\frc{\p \v{\mathsf j}_i}{\p \mathsf q_l} = 
	\sum_k \mathsf C^{lk}\bra Q_k,\v j_i\ket^{\rm c}_{\underline\beta}.
\eeq
We note that the flux Jacobian vector satisfies
\beq\label{ACsymm}
	\v{\mathsf A}\mathsf C = \mathsf C \v{\mathsf A}^{\rm T}
\eeq
where the transpose ${}^{\rm T}$ applies on each matrix $\mathsf A_\mu$ in the vector of matrices. This is shown in various contexts and levels of generality in \cite{toth2003onsager,grisi2011current,spohn2014nonlinear,castro2016emergent,de2019diffusion,karevski2019charge,doyon2020lecture,doyon2021free}. This implies that each element of the flux Jacobian vector has real eigenvalues, because the matrix $\sqrt{\mathsf C}^{-1}\mathsf A_\mu \sqrt{\mathsf C}$, obtained by a similarity transformation, is real and symmetric. It is convenient to define
\beq\label{Ap}
	\mathsf A(\h{\v p}) := \h{\v p}\cdot \v{\mathsf A}
\eeq
where $\h{\v p} = \v p/|\v p|\in\mathbb S^{d-1}$ is the unit direction vector associated to $\v p\in\R^d$. In particular
\beq
	{\rm spec}\,\mathsf A(\h{\v p}) = \{\mathsf v_I(\h{\v p})\}_I\subseteq \R
\eeq
where we use the capital letter $I$ to indicate that it runs over the {\em normal mode basis}. Clearly,
\beq
	\{\mathsf v_I(-\h{\v p})\}_I = \{-\mathsf v_I(\h{\v p})\}_I.
\eeq
We will see that the quantities $\mathsf v_I(\h{\v p})$ have the interpretation as the hydrodynamic velocities, or generalised sound velocities, for wave propagation in the direction $\h{\v p}$. Further, with the orthogonal matrix $M(\h{\v p})$ diagonalising the symmetric matrix $\sqrt{\mathsf C}^{-1}\mathsf A(\h{\v p}) \sqrt{\mathsf C}$, that is
\beq
	M(\h{\v p})\sqrt{\mathsf C}^{-1}\mathsf A(\h{\v p}) \sqrt{\mathsf C}M(\h{\v p})^{\rm T} = {\rm diag}\,(\mathsf v_I(\h{\v p}))_I,
\eeq
 the transformation to the normal modes can be chosen as $\mathsf R(\h{\v p}) = M(\h{\v p})\sqrt{\mathsf C}^{-1}$:
\beq\label{normalmodes}
	q_I^{(\h{\v p})} =\sum_j \mathsf R(\h{\v p})_I^{~j}q_j,\quad
	\mathsf R(\h{\v p}) \mathsf  A(\h{\v p}) \mathsf  R(\h{\v p})^{-1} = {\rm diag}\,(\mathsf v_I(\h{\v p}))_I,\quad \mathsf R(\h{\v p}) \mathsf C \mathsf R(\h{\v p})^{\rm T} = {\bf 1}
\eeq
for every $\h{\v p}\in\mathbb S^{d-1}$. Note how the normal modes $q_I^{(\h{\v p})}$ depend not only on the normal mode index $I$, but on the choice of the spatial direction $\h{\v p}$. In $d=1$, the $\h{\v p}$ dependence disappears,
\beq\label{normalmodes1d}
	q_I =\sum_j \mathsf R_I^{~j}q_j,\quad
	\mathsf R \mathsf  A \mathsf  R^{-1} = {\rm diag}\,(\mathsf v_I)_I,\quad \mathsf R \mathsf C \mathsf R^{\rm T} = {\bf 1}.
\eeq

Another quantity of interest is the flux Hessian vector $\v{\mathsf A}_i^{~jk} = ({\mathsf A_{1}}_i^{~jk},\ldots,{\mathsf A_{d}}_i^{~jk})$, which we denote by the same letter, but which has three indices instead of two, besides the vectorial index. This is the vector of 3-tensors given by
\beq\label{Atensor}
	\v{\mathsf A}_i^{~jk}
	:=\frc{\p}{\p\mathsf q_k}\v{\mathsf A}_i^{~j} = \frc{\p^2 \v{\mathsf j}_i}{\p \mathsf q_k\p \mathsf q_l} = 
	\sum_{mn} \mathsf C^{km}\mathsf C^{kn}\bra Q_m,Q_n,\v j_i^-\ket^{\rm c}_{\underline\beta}.
\eeq
Clearly, $\v{\mathsf A}_i^{~jk} = \v{\mathsf A}_i^{~kj}$. Further, using \eqref{ACsymm}, we obtain another symmetry relation for this quantity:
\beq
	\frc{\p \v{\mathsf A}}{\p\mathsf q_k} = 
	\frc{\p}{\p\mathsf q_k}(\mathsf C\v{\mathsf A}^{\rm T}\mathsf C^{-1})
	=
	[\frc{\p \mathsf C}{\p\mathsf q_k} \mathsf C^{-1}, \v{\mathsf A}]
	+\mathsf C\frc{\p\v{\mathsf A}^{\rm T}}{\p\mathsf q_k}\mathsf C^{-1}
\eeq
where
\beq
	\frc{\p}{\p\mathsf q_k} \mathsf C_{ij}
	= \sum_l\mathsf C^{kl} \mathsf C_{ijl},\quad
	\mathsf C_{ijl} = \bra Q_i,Q_j,q_l\ket^{\rm c}_{\underline\beta}\ \mbox{ (fully symmetric on $i,j,l$)}.
\eeq
Hence we have
\beq
	\frc{\p\v{\mathsf A}}{\p\mathsf q_k} \mathsf C
	-
	\mathsf C \frc{\p\v{\mathsf A}^{\rm T}}{\p\mathsf q_k}
	=[\frc{\p \mathsf C}{\p\mathsf q_k} \mathsf C^{-1}, \v{\mathsf A}].
\eeq
Taking the component along $\h{\v p}$ and going to its normal modes,
\beq\label{Atensornormal}
	\mathsf A_{I}^{~JK}(\h{\v p}) = \sum_{ijk}\mathsf R(\h{\v p})_I^{~i}\,\h{\v p}\cdot \frc{\p^2 \v{\mathsf j}_i}{\p \mathsf q_j\p\mathsf q_k} (\mathsf R(\h{\v p})^{-1})_j^{~J} (\mathsf R(\h{\v p})^{-1})_k^{~K} = \bra Q_J^{(\h{\v p})},Q_K^{(\h{\v p})},{\h{\v p}\cdot \v j_I^{(\h{\v p})}}^-\ket^{\rm c}
\eeq
and similarly for $\mathsf C_{IJK}(\h{\v p})$, where $\h{\v p}\cdot \v j_I^{(\h{\v p})} =\sum_k \mathsf R(\h{\v p})_I^{~k}\, \h{\v p}\cdot \v j_k$, we obtain
\beq\label{sym3pt}
	\mathsf A_{I}^{~JK}(\h{\v p})
	-
	\mathsf A_{J}^{~IK}(\h{\v p})
	= (\mathsf v_J(\h{\v p}) - \mathsf v_I(\h{\v p}))
	\mathsf C_{IJK}(\h{\v p}).
\eeq
In $d=1$, the $\h{\v p}$ dependence disappears and we have
\beq\label{Atensornormal1d}
	\mathsf A_{I}^{~JK} = \sum_{ijk}\mathsf R_I^{~i}\frc{\p^2 \mathsf j_i}{\p \mathsf q_j\p\mathsf q_k} (\mathsf R^{-1})_j^{~J} (\mathsf R^{-1})_k^{~K} = \bra Q_J,Q_K,j_I^-\ket^{\rm c}.
\eeq
and
\beq\label{sym3pt1d}
	\mathsf A_{I}^{~JK}
	-
	\mathsf A_{J}^{~IK}
	= (\mathsf v_J - \mathsf v_I)
	\mathsf C_{IJK}.
\eeq

Finally, in many physically relevant examples there is Galilean or Lorentz invariance. We will not discuss these in details, but we mention that in general, with invariance under spatial transformations $\Lambda_\mu^{~\nu}$, the flux Jacobian vector forms a projective representation,
\beq
	\sum_{\nu=1}^d \Lambda_\mu^{~\nu}\mathsf A_\nu
	=
	\mathsf K(\Lambda) \mathsf A_\mu \mathsf K(\Lambda)^{-1}.
\eeq
For instance, rotation invariance (part of Galilean and Lorentz invariance) implies that the set of hydrodynamic velocities is independent of the direction,
\beq
	\{\mathsf v_I(\h{\v p})\}_I = \{\mathsf v_I\}_I.
\eeq

\section{Proof of the projection formula}\label{appproof}

We consider the context of Sec.~\ref{sectheorem}.

\subsection{Malyshev’s formula: partial moment to cumulant}

Here denote, for lightness of notation, the state as $\bra \cdots\ket$ instead of $\dbra\cdots\dket_\ell$, and admit the case of the algebra $\mathcal A$ being non-commutative. In order to emphasise this, we adapt the notation and use $A$ and $A_i$'s, instead of $o$ and $o_i$'s, for generic elements of $\mathcal A$.

We denote by $P(n)$ the set of partitions of $\{1,2,\ldots,n\}$, and more generally $P(S)$ for the set of partitions of the set $S$. Given $\Gamma\in P(S)$, we define the set of partially connected partitions of $S$ with respect to $\Gamma$, as those in which a {\em connectedness condition} holds: no strict subset of $\Gamma$ is not connected by a partition to its complement:
\beq
	P^{\rm c}(\Gamma) = \Big\{\Pi\in P(S)\ :\  \not\exists\  \gamma\subset \Gamma,\, \pi\subset\Pi \ |\ \bigcup \gamma = \bigcup \pi\Big\}.
\eeq
That is, the least upper bound (under the ``finer than'' $\leq$ partial order) of $\Gamma$ and $\Pi$ is the trivial, single-element partition: $\Gamma\vee\Pi = \hat 1$. For an ordered set $V=(i_1,i_2,\ldots,i_{|V|})$, we denote $\bra A_i:i\in V\ket^{\rm c} := \bra A_{i_1},\ldots,A_{i_{|V|}}\ket^{\rm c}$ the corresponding cumulant (connected correlation function); we must keep track of the ordering in order to admit the case where the algebra of observables is not commutative. Recall the moment-cumulant formula, relating averages of products of observables to cumulants:
\beq\label{momentcumulant}
	\bra A_1\cdots A_n\ket = \sum_{\Pi\in P(n)}\prod_{V\in\Pi}
	\bra A_i:i\in V\ket^{\rm c}
\eeq
where every part $V$ in the partition $\Pi$ inherits its order from $(1,2,\ldots,n)$. Here and below, for an ordered set $V=(j_1,j_2,\ldots,j_{|V|})$ of positive integers, we denote $\bra A_j:j\in V\ket^{\rm c}_\ell := \bra A_{j_1},\ldots,A_{j_{|V|}}\ket^{\rm c}_\ell$ the corresponding cumulant; we must keep track of the ordering in the non-commutative case. 

Consider a contiguous partition $\Gamma\in P(n)$, that is $\Gamma=\{g_k\}_{k=1}^m$ with $g_k=\{i_k,i_k+1,\ldots,i_k+|g_k|-1\}$, $m\leq n$, and consider for every $k$ the product $A_{g_k} = A_{i_k}A_{i_k+1}\cdots A_{i_k+|g_k|-1}$. The following Lemma for {\em partial cumulants}, which generalises the moment-cumulant formula, is known as Malyshev’s formula \cite{malyshev1980cluster}, see also the proof in \cite[Prop 3.1]{peccati2008moments}. We provide a proof for completeness (in the more general setup of non-commutative algebra.
\begin{lemma}\label{lemformula}
For $g_1,\ldots,g_m$ as above:
\beq\label{formula}
	\bra A_{g_1},\ldots,A_{g_m}\ket^{\rm c} = \sum_{\Pi\in P^{\rm c}(\{g_1,\ldots,g_m\})}\prod_{V\in\Pi}
	\bra A_i:i\in V\ket^{\rm c}
\eeq
where the sum is over partially connected partitions of $\{1,\ldots,n\}$ with respect to $\Gamma = \{g_k\}_{k=1}^m$, and every part $V$ in the partition $\Pi$ inherits its order from $(1,2,\ldots,n)$.
\end{lemma}
\proof
Consider a contiguous partition $\Gamma=\{g_1,\ldots,g_r\}\in P(n)$ as above. Because the partition is contiguous, it inherits an ordering, which makes $\Gamma$ into an ordered set $(g_1,g_2,\ldots g_r)$. By associativity of the product of observables, we may write the moment-cumulant formula, instead of \eqref{momentcumulant}, as
\beq\label{momentgcumulant}
	\bra A_1\cdots A_n\ket = \bra A_{g_1}\cdots A_{g_r}\ket =
	\sum_{\Omega\in P(\Gamma)}\prod_{G\in\Omega}
	\bra A_{g}:g\in G\ket^{\rm c}
\eeq
where every part $G$ of the partition $\Omega$ inherits its order from $\Gamma$. We compare \eqref{momentcumulant} with \eqref{momentgcumulant} and proceed by induction. We note that \eqref{formula} holds for $m=1$ as it reproduces \eqref{momentcumulant}. Assume that it holds for all $m\leq r-1$ for some $r\geq 2$, for all $n\geq 1$, and for all contiguous partitions $\{g_k\}_{k=1}^m$. Then we will show that it holds for $m=r$. For the rest of the proof, we fix $n$, $r$ and a contiguous, ordered partition $\Gamma = (g_k)_{k=1}^r$.

For every $\Pi\in P(n)$, consider the least upper bound  $\Pi\vee \Gamma$ of the set of partitions $\{\Pi, \Gamma\}$. Because $\Gamma$ is finer than $\Pi\vee \Gamma$, this defines a partition $\Omega\in P(\Gamma)$: the one such that for all $G\in\Omega$, we have $\bigcup G \in \Pi\vee \Gamma$. We note that this is the unique $\Omega\in P(\Gamma)$ such that for all $G\in\Omega$, there exists $\pi\subseteq \Pi$ with $\pi\in P^{\rm c}(G)$. Indeed, suppose $\Xi$ is such that $\Pi\leq \Xi,\,\Gamma\leq \Xi$, and define $\Omega\in P(\Gamma)$ with respect to $\Xi$. As for every $G\in\Omega$ we have $\bigcup G\in \Xi$, then for every $G\in\Omega$ there exists $\pi\subseteq \Pi$ such that $\bigcup G = \bigcup \pi$ (because $\Pi$ is finer than $\Xi$), thus $\pi\in P(\bigcup G)$. But also, there is a finer partition than $\Xi$, that still upper-bounds $\{\Pi,\Gamma\}$, if and only if there is at least one such $G\in\Omega$ such that $\exists \gamma\subset G, \pi'\subset\pi\,|\,\bigcup\gamma=\bigcup\pi'$. Thus for the finest $\Xi$, we have $\pi \in P^{\rm c}(G)\subset P(\bigcup G)$. We denote the resulting $\Omega = \xi(\Pi)$ (recall that $\Gamma$ is fixed, so we do not display the dependence on $\Gamma$). 
%Given any $\Omega\in P(\Gamma)$ one may define a partition of $\{1,\ldots,n\}$ as $\rho(\Omega):= \{\bigcup G:G\in\Omega\}\in P(n)$, and we note that the map $\rho:P(\Gamma)\to P(n)$ preserves the ``finer than" ($\leq$) partial order. Because of the partial ordering, for every partition $\Pi\in P(n)$ there is a unique minimal $\Omega \in P(\Gamma)$ such that $\Pi\leq\rho(\Omega)$. We denote this as $\xi(\Pi) = {\rm min}(\Omega: \Pi\leq\rho(\Omega))$; this is the finest $\Omega$ that partitions $\Gamma$ into subsets whose unions contain full parts only. We note that $\xi(\Pi)$ is the unique $\Omega$ such that for all $G\in\Omega$, there exists $\pi\subset \Pi$ such that $\bigcup G = \bigcup \pi$ (that is $\Pi\leq \rho(\Omega)$), and $\pi\in P^{\rm c}(G)$ (minimality). 
By the uniqueness of $\Omega$ thus defined, we have that, given any $\Omega\in P(\Gamma)$, the set of all $\Pi\in P(n)$ such that $\Omega = \xi(\Pi)$ is
\beq\label{xiinv}
	\xi^{-1}(\Omega) =
	\Bigg\{\bigcup_{G\in\Omega} \Pi_G \ :\  \Pi_G \in  P^{\rm c}(G)\;\forall\; G\in\Omega\Bigg\},
\eeq
and also, we have the important decomposition
\beq\label{decompo}
	P(n) = \bigcup_{\Omega\in P(\Gamma)} \xi^{-1}(\Omega).
\eeq

On the right-hand side of \eqref{momentgcumulant} consider all  partitions $\Omega$ with $|\Omega|\geq 2$. Then for every such term, every factor in the product has $|G|\leq r-1$, so we can use the induction hypothesis (noting that for every $G$, the ordered set $(g:g\in G)$ is a contiguous partition of the ordered set $\bigcup G$). Thus for these terms we have, using \eqref{xiinv},
\beq
	\sum_{\Omega\in P(\Gamma)\atop |\Omega|\geq 2}
	\prod_{G\in\Omega}
	\Bigg(
	\sum_{\Pi\in P^{\rm c}(G)}
	\prod_{V\in\Pi}
	\bra A_i:i\in V\ket^{\rm c}
	\Bigg)
	=
	\sum_{\Omega\in P(\Gamma)\atop |\Omega|\geq 2}
	\sum_{\Pi\in \xi^{-1}(\Omega)}\;
	\prod_{V\in\Pi}
	\bra A_i:i\in V\ket^{\rm c}.
\eeq
There is a unique partition $\Omega\in P(\Gamma)$ with $|\Omega|=1$: it is $\Omega = \{\Gamma\}$. Therefore, comparing with \eqref{momentcumulant}, we obtain, using \eqref{decompo}
\beqa
	\bra A_{g_1},\ldots,A_{g_r}\ket^{\rm c}
	&=&
	\Bigg(\sum_{\Pi\in P(n)} - \sum_{\Omega\in P(\Gamma)\atop |\Omega|\geq 2}
	\sum_{\Pi\in \xi^{-1}(\Omega)}\Bigg)\prod_{V\in\Pi}
	\bra A_i:i\in V\ket^{\rm c}\n
	&=&\sum_{\Pi\in \xi^{-1}(\{\Gamma\})}\prod_{V\in\Pi}
	\bra A_i:i\in V\ket^{\rm c}\n
	&=& \sum_{\Pi\in P^{\rm c}(\Gamma)} \prod_{V\in\Pi}
	\bra A_i:i\in V\ket^{\rm c}
\eeqa
which completes the proof.
\eproof

\subsection{Hydrodynamic projection}\label{ssectproj}

We recall that we denote by $C(n)$ the set of {\em connected covers} of $\{1,\ldots,n\}$. For the purpose of this proof, however, a connected cover is a cover $\Upsilon = \{V_1,V_2,\ldots,V_m\}$, considered as a set of patches $V_k\subseteq \{1,\ldots,n\}$ {\em with multiplicities} (that is, some sets $V_i$ may appear many times). Usually one does not consider multiplicities, but considering multiplicities is convenient in the proof below. Importantly, Lemma \ref{lemmaxindex} still holds on the set of such covers. Each patch is a set (without multiplicities -- that is, each patch is an ordinary set, it does not have repeated elements), such that $\cup_k V_k = \{1,\ldots,n\}$, and, as mentioned before, such that the connectedness condition holds: there is no subset of indices $K$ with $\cup_{i\in K} V_i \cap \cup_{i\not\in K}V_i = \emptyset$. That is, no subset of patches whose union is disconnected from the complementary subset of patches. Recall that for a cover $\Upsilon\in C(n)$, we denote by $m_\Upsilon(k)$ the multiplicity of the element $k\in \{1,\ldots,n\}$ within the cover $\Upsilon$, that is $m_\Upsilon(k) = |\{V\in\Upsilon: k\in V\}|$.

We now prove Theorem \ref{theomain}.

\proof The existence of \eqref{So1} and the statement \eqref{onept} follows from \eqref{Soproofq} and the moment-cumulant formula \eqref{momentcumulantmain}: in this formula, the single dominant term as $\ell\to\infty$ is that which is fully factorised (if the state comes from a probability measure, one may simply use the Bienaym\'e-Chebyshev inequality).

We note that \eqref{generalprojectionproof} is multilinear in the observable (algebra element) $\mathsf o(\underline{ q})$. Hence, if \eqref{generalprojectionproof} holds for all variables of the monomial, centered form:
\beq\label{centeredo0}
	\mathsf o(\underline{ q}) = \delta q_{a_{1}}\cdots \delta q_{a_{r}}
\eeq
with $\delta q_a = q_a - \dbra q_a(z)\dket\,\1$ and $\dbra q_a(z)\dket=\lim_{\ell\to\infty} \dbra q_a\dket_\ell$, then it holds for all observables.

Consider
\beq\label{centeredo}
	\mathsf o_k(\underline{ q}) = \delta q_{a_{k1}}\cdots \delta q_{a_{kr_k}},\quad k=1,2,\ldots,n
\eeq
for algebra element indices $a_{kl}\in \alpha$ parametrised by two indices $k,l$, where the range $l=1,\ldots,r_k$ may depend on $k$. We use \eqref{formula} in order to evaluate
\[
	\dbra \mathsf o_1(\underline q),\ldots,\mathsf o_n(\underline q)\dket^{\rm c}_\ell.
\]
In \eqref{formula}, the state $\bra\cdots\ket$ is replaced by the state $\dbra\cdots\dket_\ell$, every observable $A_j$ is now $A_{kl}$, which has two indices, and is set to $A_{kl} = \delta q_{a_{kl}}$, and the parts $g_k$ in the contiguous partition $\Gamma$ are taken such that $A_{g_k} = \mathsf o_k(\underline q)$. Also, we replace $m$ by $n$ in \eqref{formula} in order to agree with the choice of variable made in \eqref{generalprojectionproof} (that is, $k\in\{1,\ldots,n\}$). We take $n\geq 2$.

Every partially connected partition $\Pi$ in the sum in \eqref{formula} gives rise to a connected cover $\Upsilon\in C(n)$ of $\{1,\ldots,n\}$. This is done as follows. We associate to every part
\[
	W=\{(k_1,l_1),\ldots,(k_{|W|},l_{|W|})\}\in \Pi
\]
in the partition, the patch $V^\sharp = \{k_1,\ldots,k_{|V|}\}\in\Upsilon$. This is the set associated to the sequence $V=V(W):=(k_1,\ldots,k_{|V|})$ (with $|V|=|W|$), where we discount possible multiplicities -- a particular value may be repeated in the sequence, $k_s=k_{s'}$ for some $s\neq s'$ (in this case, we necessarily have $l_s\neq l_{s'}$), but if this happens, we put this value only once in $V^\sharp$. For definiteness, the sequence is specified, say, by the non-decreasing order, $k_s\leq k_{s+1}$. The fact that this gives rise to a connected cover, is because every partition is such that the set of patches covers $\{1,\ldots,n\}$ by definition of a partition, that is $\cup_{W\in\Pi} V(W)^\sharp = \{1,\ldots,n\}$; and the connectedness condition of the partition implies the connectedness condition of the cover. As per \eqref{formula}, we associate to every part $W$ the factor
\[
	\dbra \delta q_{a_{k_1l_1}} ,\ldots, \delta q_{a_{k_{|V|}l_{|V|}}}\dket^{\rm c}_\ell,
\]
over which we take the product for a given partition. Thus, we have to sum such products, over all connected covers $\Upsilon$ that are associated to connected partitions.

In this sum over covers, a given cover $\Upsilon$ may occur multiple times, as any given cover may be associated to many partitions, because of the freedom on the $l_s$'s. Further, $\Upsilon$ is itself a set (of patches) with multiplicities, as different parts $W$ in a partition may lead to the same patch $V(W)^\sharp$. Given a patch $V^\sharp\in\Upsilon$, the choice of sequences $V=(k_1,\ldots,k_{|V|})$ and $L=(l_1,\ldots,l_{|V|})$ depends on the copy of $V^\sharp$ if $V^\sharp$ has multiplicity greater than 1 in $\Upsilon$. $V$ and $L$ also depend on the copy of $\Upsilon$ if this given cover occurs multiple times.

So we may represent the sum over partially connected partitions in \eqref{formula} as a sum of terms, each of the form of a product over all patches $V^\sharp\in\Upsilon$,
\beq\label{produpsilon}
	\prod_{V^{\sharp}\in\Upsilon:\atop V=(k_1,\ldots,k_{|V|}),\,L=
	(l_1,\ldots,l_{|V|})}\dbra \delta q_{a_{k_1l_1}} ,\ldots, \delta q_{a_{k_{|V|}l_{|V|}}}\dket^{\rm c}_\ell,
\eeq
for some $\Upsilon\in C(n)$, where $V$ and $L$ depend on $V^\sharp$, and in particular on the copy of $V^\sharp$ in $\Upsilon$ if $V^\sharp$ has multiplicity higher than 1.

In order to determine which $V$'s and $L$'s occur, we use the statement \eqref{Soproofq}, expressed as $\dbra q_{a_1},\ldots,q_{a_n}\dket^{\rm c}_\ell = \dbra \delta q_{a_1},\ldots,\delta q_{a_n}\dket^{\rm c}_\ell =\mathcal O(\ell^{1-n})$. In this sense, the term \eqref{produpsilon} is
\[
\mathcal O(\ell^{|\Upsilon| - \sum_{V^{\sharp}\in\Upsilon} (|V^\sharp|+ M_V)})\]
where $|V^\sharp|+ M_V = |V|$, that is, $M_V = \sum_{k\in V^{\sharp}} (M(k)-1)$ and $M(k) = |\{s:k_s=k, V= (k_1,\ldots,k_{|V|})\}|$ is the multiplicity of $k\in V$.

Let us concentrate on the ``leading terms'' \eqref{produpsilon} which are such that $M_V=0$ for all $V^\sharp\in\Upsilon$ -- that is, $V=V^\sharp$ (if discounting the ordering in $V$) -- and such that the index $\iota(\Upsilon)= \sum_{V^{\sharp}\in\Upsilon} |V^\sharp|-|\Upsilon|$ of $\Upsilon$ takes value $\iota(\Upsilon)=n-1$. By Lemma \ref{lemmaxindex}, any other terms will be $\mathcal O(\ell^{u})$ for $u<1-n$. Therefore, we have shown that $\mathsf S_{o_1,\ldots,o_n}$, defined as the limit Eq.~\eqref{Soproofo}, exists for all $n\geq 2$ (the case $n=1$ was already shown), and that these other terms give zero contribution to this limit.

For the leading terms, multiplicities of elements in $V$ are all $M(k)=1$. Hence any singleton $V\in\Upsilon\in C(n)$ is associated with a factor $\dbra \delta q_{a_{kl}}\dket_\ell$ which vanishes in the limit, $\lim_{\ell\to\infty} \dbra \delta q_{a_{kl}}\dket_\ell=0$, by centering in \eqref{centeredo}. Thus, we are left with connected covers which have no singletons, and with index $n-1$. These are the minimal connected covers $C^{\rm min}(n)$.

Taking the limit, there is some subset $C^\sharp\subseteq C^{\rm min}(n)$ such that, adjoining appropriate multiplicities $C^\sharp\to C$, we have
\beq\label{Sointermediate}
	\mathsf S_{o_1,\ldots,o_n}
	= \sum_{\Upsilon\in C}
	\prod_{V=\{k_1,\ldots,k_{|V|}\}\in\Upsilon\atop
	(l_1,\ldots,l_{|V|})} \mathsf S_{a_{k_1l_1},\ldots, a_{k_{|V|}l_{|V|}}}.
\eeq
Because the cover $\Upsilon$ is minimal, its elements have no multiplicities: each set $V\in\Upsilon$ is present only once, and hence $(l_1,\ldots,l_{|V|})$ is a sequence that is a function of $V$. But a cover $\Upsilon\in C$ may be present multiple times (may have multiplicity greater than 1), hence give rise to many terms, and the function $V\mapsto L=(l_1,\ldots,l_{|V|})$ depends on the copy of the cover taken.

Finally, we have to determine the exact set of partitions associated to each cover $\Upsilon\in C^{\rm min}(n)$. For a given patch $V=\{k_1,\ldots,k_{|V|}\}\in\Upsilon$, the set of all possible associated parts, in the partition $\Gamma$ giving rise to this cover, are $W=\{(k_1,l_1),\ldots,(k_{|V|},l_{|V|})\}$ with all choices of $L=(l_1,\ldots, l_{|V|})$ such that $l_s \in \{1,\ldots,r_{k_j}\}$, with the sole constraint that any element $k$ common to a certain number of patches, must come, on the different patches, with different values of $l$'s. These certainly cover all the possible partitions $\Gamma$, and, importantly, do not overcount them, because, as we have established, no two $V$'s are equal (so there are no two choices of sequences $L$ that would just interchange two such $V$'s).

Note that the sequence $L=(l_1,\ldots, l_{|V|})$ is unambiguous from $W$: we have the non-decreasing condition $k_s\leq k_{s+1}$, and as $V$ is a set, this becomes the strictly increasing condition $k_s< k_{s+1}$, thus uniquely fixing what we mean by the sequence $l_1,\ldots,l_{|V|}$. This gives, for every $V$, a function $l^V:k\in V\to \{1,\ldots,r_k\}$ with the constraint $l^V(k)\neq l^{V'}(k)$ for every $V,V'\in\Upsilon_k$, where we recall that $\Upsilon_k = \{V\in \Upsilon: k\in V\}$.

In order to obtain the final formula, we must translate this description, which is in terms of $(k,l)$, into a description in terms of algebra element index $a\in \alpha$. This is done by using derivatives of $\mathsf o_k(\underline{\mathsf q})$ evaluated at $\mathsf q_a = \dbra q_a\dket$. Indeed, by the centering \eqref{centeredo0}, these derivatives vanish unless all factors present have been differentiated, and this implements the constraint above. However, we must be careful to count appropriately, because indices $a_{kl}$ may repeat. We do this as follows.

Given $k$, consider the equivalence relation $l\sim_k l'$ defined by $a_{kl}=a_{kl'}$. We define the symmetry factor $s_k$ as the following product over distinct equivalence classes $[l]_k = \{l':l'\sim_k l\}$:
\beq
	s_k = \prod_{[l]_k} |[l]_k|!\,.
\eeq
Recall that $m_\Upsilon(k) = |\Upsilon_k|$ is the number of times $k$ is included within a patch $V\in\Upsilon$. Given $k$, let $a^V_k\in \alpha$ for all $V\in\Upsilon_k$. Then
\beq\label{derconstraint}
	\frc{\p^{m_\Upsilon(k)}\mathsf o_k(\underline{\mathsf q})}{\prod_{V\in\Upsilon_k} \p \mathsf q_{a^V_k}}\Bigg|_{\underline{\mathsf q} = \dbra \underline{q}\dket}
	=
	\lt\{\ba{ll}
	s_k & (\{a_k^V:V\in\Upsilon_k\} = \{a_{k1},\ldots,a_{kr_k}\})
	\\
	0 & \mbox{(otherwise).}
	\ea\rt.
\eeq
Note how the factor $s_k$ arises because of multiple differentiations with respect to a given, repeated index. For fixed $k$, given $a_k^V:V\in\Upsilon_k$'s such that $\{a_k^V:V\in\Upsilon_k\} = \{a_{k1},\ldots,a_{kr_k}\}$, the condition $a_k^V = a_{k\,l^V(k)}\,\forall\;V\in\Upsilon_k$ gives rise to $s_k$ choices of $l^V(k)$ with the constraint above. As, for each such choice made for every $k$, the quantity
\[
	\prod_{V=\{k_1,\ldots,k_{|V|}\}\in\Upsilon} \mathsf S_{a_{k_1\,l^V(k_1)},\ldots, a_{k_{|V|}\,l^V(k_{|V|})}}
	=
	\prod_{V=\{k_1,\ldots,k_{|V|}\}\in\Upsilon} \mathsf S_{a_{k_1}^V,\ldots, a_{k_{|V|}}^V}
\]
is the same, the sum over these choices of $l^V(k)$ gives that quantity times the factor
\[
	\prod_{k=1}^n \frc{\p^{m_\Upsilon(k)}\mathsf o_k}{\prod_{V\in\Upsilon_k} \p \mathsf q_{a^V_k}}\Bigg|_{\underline{\mathsf q} = \dbra \underline{q}\dket}.
\]
Summing over all values of $a_k^V\in \alpha$, by \eqref{derconstraint} this gives the sum over all possible $l^V(k)$. Therefore, the sum \eqref{Sointermediate}, accounting for the multiplicities of $\Upsilon$ coming from the choices of $l^V(k)$, gives \eqref{generalprojectionproof}. This completes the proof.
\eproof

\section{Three-point amplitude solution and its time-translation invariance} \label{app3pt}

We consider the solution, Eqs.~\eqref{solSIJK}, \eqref{solSIJK2}, for the three-point amplitude $\mathsf S_{IJK}$. We show that this is indeed a solution to \eqref{eqSIJK} with \eqref{initSIJK}, and that it is invariant under time translation.

We first note that the cases $v=0,\,v'\neq0$ and $v\neq 0,\,v'=0$ are obtained from the careful limit $v\to0$ and $v'\to0$, respectively, from the case $v\neq 0,\,v'\neq0$. Take for instance $v=0,\,v'\neq 0$. The limit on the first term in the parenthesis of the top line of \eqref{solSIJK2} is immediate and gives the second term in the parenthesis of the second line of \eqref{solSIJK2}. For the second term of the top line of \eqref{solSIJK2}, one has to be careful as a naive limit would give $|v|$ times a product which tends to $s(u)\delta'(u)$, which is ill-defined. Instead, this term must be written
\beqa
	- |v|\,\delta'(v'u-vu') s(u+vt) &=&
	-\frc{|v|}{|v'|v'}\,\delta'(u-vu'/v') s(u+vt)
	\n &=&
	-\frc{|v|}{|v'|v'}\,\Big(\p_u\big(\delta(u-vu'/v') s(u+vt)\big)
	- \delta(u-vu'/v') \delta(u+vt)\Big)
	\n &=&
	-\frc{|v|}{|v'|v'}\,\Big(\delta'(u-vu'/v') s(vu'/v'+vt)
	- \frc{|v'|}{|v|} \delta(u-vu'/v') \delta(u'+v't)\Big)\n
	&\stackrel{v\to0}=&
	\frc1{v'}\delta(u)\delta(u'+v't)
\eeqa
where the limit $v\to0$ is well defined on each term because each is a product of functionals for different variables. Obtaining the case $v=0,\,v'=0$ from the other ones does not appear to be well justified, so this case must be treated separately. Hence, we will consider the cases $v\neq0,\,v'\neq 0$, and $v=0,\,v'=0$ only.

\subsection{Proof that it is a solution}

The fact that  \eqref{solSIJK} satisfies the initial condition \eqref{initSIJK} is immediate, as only the first term in \eqref{solSIJK} remains at $t_1=t_2=t_3=0$.

To show that it is indeed a solution to \eqref{eqSIJK}, we remark that the all variables $u_a = x_a-\mathsf v_a t_a$, $a=1,2,3$ are invariant under all tansport derivatives $\p_{t_b}+ \mathsf v_b \p_{x_b}$, $b=1,2,3$ (whether $a=b$ or not). Therefore in \eqref{eqSIJK}, the only nonzero contribution from the transport derivative $\p_{t_1}+ \mathsf v_I \p_{x_1}$ on the solution \eqref{solSIJK}, comes trom the time derivative $\p_{t_1}$ applied on the explicit time dependence of the single auxiliary function $b_{t_1}(u_{12},u_{13},\mathsf v_{IJ},\mathsf v_{IK})$; similarly for the equations obtained by cyclic permutations. We show that this contribution exactly cancels the second term (source term) in \eqref{eqSIJK}.

The case $v=0,\,v'=0$ is immediate as time appears only as an overall factor, and as all velocities are equal $\mathsf v_I=\mathsf v_J = \mathsf v_K$, we have $u=x_{12}-\mathsf \mathsf v_J t_{12}$, $u'=x_{13}-\mathsf \mathsf v_K t_{13}$; thus $\p_t b_t(u,u',v,v') = \p_{x_1}\big(\delta(x_{12}-\mathsf v_J t_{12}) \delta(x_{13}-\mathsf v_K t_{13})\big)$, which shows the cancellation. In the case $v\neq0,\,v'\neq 0$, the derivative gives
\beqa
	\p_t b_t(u,u',v,v')
	&=&
	\delta'(v'u-vu') \Big(|v'|v' \delta(u'+v't) - |v|v \delta(u+vt)\Big)
	\n
	&=&
	\delta'(u-vu'/v') \delta(u'+v't) + \delta'(u'-v'u/v) \delta(u+vt)
	\n
	&=&
	\delta'(u+vt) \delta(u'+v't) + \delta'(u'+v't) \delta(u+vt)
\eeqa
and using $u + v t = u_{12} +\mathsf v_{IJ} t_1 = x_{12} -\mathsf v_I t_1 + \mathsf v_J t_2 + \mathsf v_{IJ} t_1 = x_{12} + \mathsf v_J t_{12}$ and $u' + v' t = u_{14} +\mathsf v_{IK} t_1 = x_{13} + \mathsf v_K t_{13}$ this equals $\p_{x_1}\big(\delta(x_{12}-\mathsf v_J t_{12}) \delta(x_{13}-\mathsf v_K t_{13})\big)$, which shows again the cancellation.

\subsection{Proof that it is time-translation invariance}\label{app3pttime}

The solution \eqref{solSIJK} is in principle unique. However, its time-translation invariance, which should hold as the state $\bra\cdots\ket$ is stationary, is not manifest. We now show with $\mathsf S_{IJK}(x_1,t_1;x_2,t_2;x_3,t_3)$ given by \eqref{solSIJK}, we have
\beq
	\mathsf S_{IJK}(x_1,t_1+r;x_2,t_2+r;x_3,t_3+r)
	=
	\mathsf S_{IJK}(x_1,t_1;x_2,t_2;x_3,t_3)
\eeq
for all $r$. That is, we replace $t_a \to t_a + r$ (fixing $x_a$'s) in \eqref{solSIJK}, and show that the result is independent of $r$.

In the case $v=v'=0$, we note that because all velocities are equal $\mathsf v_I=\mathsf v_J = \mathsf v_K = \mathsf v$, each $u_{ab}$ is independent of $r$. There remains only the explicit time dependence in the auxiliary function $b_t(u,u',v,v')$. Because of the relation \eqref{sym3pt1d}, the 3-point coupling $\mathsf A_I^{~JK}$, for equal velocities, is completely symmetric under exchange of its indices. Hence, the $r$-dependent terms are
\beq
	-r\mathsf A_I^{~JK}\,\Big( \p_{u_1}\big(\delta(u_{12})\delta(u_{13})\big)
	+
	\p_{u_2}\big(\delta(u_{23})
	\delta(u_{21})\big)
	+
	\p_{u_3}\big(\delta(u_{31})
	\delta(u_{32})\big)
	\Big).
\eeq
We re-write the parenthesis in terms of the irreducible set of delta functions $\delta(u_{12}),\,\delta (u_{13})$ and their derivatives, and find that it vanishes:
\beq
	\delta'(u_{12})\delta(u_{13})
	+
	\delta(u_{12})\delta'(u_{13})
	-
	\delta(u_{13})\delta'(u_{12})
	-
	\delta'(u_{13})\delta(u_{12})
	= 0.
\eeq

Finally, the case $v\neq0,\,v'\neq 0$ is the most intricate, and requires a precise cancellation between the first term in \eqref{solSIJK}, which implements the initial condition, and the cyclic sum of the other terms, again along with the relation \eqref{sym3pt1d}. We note that under $t_a\to t_a+r$ we have $u_{ab} \to u_{ab} + \mathsf v_{ab} r$ and hence in $b_t(u,u',v,v')$ we have $u\to u+vr,\,u'\to u'+v'r$ so that the quantities
\beq
	v'u-vu',\ u + vt,\ u'+v' t'
\eeq
are all independent of $r$. With $\mathsf C_{IJK}=\bra Q_I,Q_J,q_K\ket^{\rm c}$, the $r$-dependent terms are the term $\mathsf C_{IJK}\delta(u_{12})\delta(u_{13})$ and the terms involving $b_0(u,u',v,v')$:
\beq
	\mathsf C_{IJK}\delta(u_{12})\delta(u_{13})
	+\Big(\mathsf A_I^{~JK}
	\delta'(\mathsf v_{IK} u_{12}-\mathsf v_{IJ}u_{13})\big(
	|\mathsf v_{IK}|s(\mathsf u_{13})
	-
	|\mathsf v_{IJ}|s(\mathsf u_{12})
	\big)+\mbox{cyclic permutations} \Big).
\eeq
We apply $\p_r$ on these, with $\p_r u_{ab} = \mathsf v_{ab}$. The first term gives
\beq\label{Cterms}
	\mathsf C_{IJK}\Big(
	\mathsf v_{IJ} \delta'(u_{12})\delta(u_{13})
	+
	\mathsf v_{IK} \delta(u_{12})\delta'(u_{13})
	\Big).
\eeq
The first of the terms in the parenthesis gives
\beqa
	\lefteqn{\mathsf A_I^{~JK}
	\delta'(\mathsf v_{IK}u_{12}-\mathsf v_{IJ}u_{13})\big(
	|\mathsf v_{IK}|\mathsf v_{IK} \delta(u_{13})
	-
	|\mathsf v_{IJ}|\mathsf v_{IJ} \delta(u_{12})\big)} &&\n
	&=&
	\mathsf A_I^{~JK}
	\delta'(u_{12}-\mathsf v_{IJ}u_{13}/\mathsf v_{IK})
	\delta(u_{13})
	+
	\mathsf A_I^{~JK}\delta'(u_{13}-\mathsf v_{IK}u_{12}/\mathsf v_{IJ})
	\delta(u_{12})\n
	&=&
	\mathsf A_I^{~JK}\p_{u_1}
	\big(\delta(u_{12})
	\delta(u_{13})\big)
\eeqa
so that the sum over permutations is
\beq
	\mathsf A_I^{~JK}\p_{u_1}
	\big(\delta(u_{12})
	\delta(u_{13})\big)
	+
	\mathsf A_J^{~KI}
	\p_{u_2}
	\big(\delta(u_{23})
	\delta(u_{21})\big)
	+
	\mathsf A_K^{~IJ}
	\p_{u_3}\big(
	\delta(u_{31})
	\delta(u_{32})
	\big).
\eeq
We re-write this in terms of the irreducible set of delta functions $\delta(u_{12}),\,\delta (u_{13})$ and their derivatives, and we find
\beq\label{Aterms}
	\Big(\mathsf A_I^{~JK} - \mathsf A_J^{~IK}\Big)
	\delta'(u_{12})\delta(u_{13})
	+
	\Big(\mathsf A_I^{~KJ} - \mathsf A_K^{~IJ}\Big)
	\delta(u_{12})\delta'(u_{13})
\eeq
where we have used symmetry under exhange of the upper indices. The relation \eqref{sym3pt1d} guarantees that the sum of \eqref{Cterms} and \eqref{Aterms} vanishes.

This shows invariance under time translations.

\end{document}